\documentclass[twocolumn,aps,epsfig,nofootinbib]{revtex4}

\usepackage{graphicx}
\usepackage{epstopdf}
\usepackage{latexsym}
\usepackage{amssymb}
\usepackage{amsmath}
\usepackage{color}
\usepackage{mathrsfs}

\usepackage[center]{subfigure}

\begin{document}

 \newcommand{\bq}{\begin{equation}}
 \newcommand{\eq}{\end{equation}}
 \newcommand{\bqn}{\begin{eqnarray}}
 \newcommand{\eqn}{\end{eqnarray}}
 \newcommand{\nb}{\nonumber}
 \newcommand{\lb}{\label}
\newcommand{\PRL}{Phys. Rev. Lett.}
\newcommand{\PL}{Phys. Lett.}
\newcommand{\PR}{Phys. Rev.}
\newcommand{\CQG}{Class. Quantum Grav.}

\title{Inflationary cosmology with  nonlinear dispersion relations}  

\author{Tao Zhu, Anzhong Wang}

 \affiliation{GCAP-CASPER, Physics Department, Baylor University, Waco, TX 76798-7316, USA \\
 Institute for Advanced Physics $\&$ Mathematics, Zhejiang University of Technology, Hangzhou, 310032, China}

 \author{Gerald Cleaver}

 \affiliation{EUCOS-CASPER, Physics Department, Baylor University, Waco, TX 76798-7316, USA}

 \author{Klaus Kirsten and Qin Sheng}

 \affiliation{GCAP-CASPER, Mathematics Department, Baylor University, Waco, TX 76798-7328, USA}

\date{\today}

\begin{abstract}

We present a technique, {\em the uniform asymptotic approximation}, to construct accurate analytical  solutions of  the linear 
perturbations of inflation  after  quantum effects of the early universe are taken into account,  for which the dispersion relations 
generically become  nonlinear.  We construct  explicitly the error  bounds associated with the approximations and then study 
them in detail. With the understanding of the errors  and the proper choice of the Liouville transformations of the differential 
equations of the perturbations, we show that  the analytical solutions describe the  exact evolution of the linear perturbations  
extremely well even only in the first-order approximations. As an application of the approximate analytical solutions, we calculate 
the power spectra and indices of scalar and tensor perturbations in the slow-roll inflation, and find that the amplitudes of the power 
spectra get modified due to the quantum effects, while the power spectrum indices remain the same as in the linear case.

\end{abstract}



\maketitle

\section{Introduction}
\renewcommand{\theequation}{1.\arabic{equation}} \setcounter{equation}{0}

The inflationary  cosmology \cite{Guth}, aside from solving several fundamental and conceptual problems, such as flatness, 
horizon, and exotic-relics, of the standard big bang cosmology, also provides a causal mechanism for generating structures 
in the universe and the spectrum of cosmic microwave background (CMB) anisotropies  \cite{InfGR}. The large scale structure 
and the CMB 
anisotropies in the universe arose from the gravitational collapse of adiabatic, Gaussian, and nearly scale-invariant primordial 
fluctuations of space-time, a key prediction of inflation models.  These are now matched to observations with unprecedented  
precision \cite{WMAP}, especially after  the recent release of  the more precise cosmological results from the Planck satellite 
\cite{PLANCK}.

Inflationary  scenario is based on the assumption that there is a period in the very early universe during which the scale of space 
is expanding at an exponentially growing rate. Consequently, the wavelengths corresponding to the present large scale structure 
in the Universe and to the measured CMB anisotropies, were exponentially stretched during inflation.  However, it was found that, 
if the inflationary period is sufficiently long to consistent with observations,
the physical wavelength of fluctuations observed at the present time may well originate with a wavelength smaller than the Planck
length at the beginning of the inflation --- the trans-Planckian issues \cite{Brandenberger1999}. Thus, questions immediately arise 
as to whether the usual predictions of the scenario still remain robust, due to the ignorance of physics in such a small  scale, 
and more interestingly, whether they could leave  imprints for future observations, even the usual predictions are indeed robust.
Such considerations have attracted a great deal of   attention  in understanding the quantum effects of the early universe  on the inflation   
and  cosmological perturbations  \cite{Martin2001,eff}.

So far, various  approaches have been proposed  to study the aforementioned effects  \cite{Brandenberger2013CQG}. One is to choose 
a different initial state rather than that of  the  Bunch-Davies vacuum \cite{Bunch-Davies1978}. This corresponds to introducing a time-like 
new physics hypersurface on which an {\em ad hoc} initial state is set. In this perspective, the quantum effects have simply been shifted to 
modifying the  initial states. Another  approach is the boundary effective field theory  \cite{eff}. Its main idea was to integrate out the high 
energy physics and then derive an effective theory which only involves observable scales.  However, the resulting effects from these two 
approaches are controlled by the initial time of inflation rather than the redshift where Planck physics is relevant. This implies a {\em priori 
assumption}  that the quantum effecters of the early universe  do not change the existing predictions,  except for some suppressed corrections.

Instead of specifying {\em ad hoc} initial states, another natural option is to replace the conventional linear dispersion relation by a nonlinear 
one. This approach was initially  applied to inflationary cosmology as a toy model  \cite{Martin2001}, motivated from the studies of the 
dependence of black hole  radiation on Planck scale physics \cite{UCJ}. Later, it was  found  \cite{HL} that it could be naturally realized   
in the framework of the Ho\v{r}ava-Lifshitz gravity  \cite{Horava,reviews}, a candidate of the ultraviolet complete theory of quantum gravity.
 The  perturbations   (of scalar, vector or tensor fields) produced during the inflationary epoch are described  by \cite{Brandenberger2013CQG},
\bqn
\lb{eom}
\mu''_k(\eta) +\left(\omega_k^2(\eta)-\frac{z''}{z}\right)\mu_k(\eta)=0,
\eqn
where   $\mu_k(\eta)$ denotes  the mode function,  a prime  the differentiation with respect to the conformal time $\eta$, and $z(\eta)$ 
depends on the background and the types of perturbations, scalar, vector or tensor.
The modified dispersion relation $\omega_k^2(\eta)$ takes the form,
\bqn
\lb{omega}
\omega^2_k(\eta) = k^2 \left[1-\hat{b}_1 \left(\frac{k}{a M_*}\right)^2+\hat{b}_2 \left(\frac{k}{a M_*}\right)^4\right],
\eqn
where
$M_*$ is the relevant energy scale of the trans-Planckian physics, $k$ is the comoving wavenumber of the mode, $\hat{b}_1$ and $\hat{b}_2$ 
are dimensionless constants, and in order to get a healthy ultraviolet limit, one requires $\hat{b}_2>0$.  When $\hat{b}_i = 0$, it reduces to that of 
general relativity (GR). 

In the slow-roll inflation, we have $a(\eta) \simeq - (1-\varepsilon)/(\eta H)$, with $H$ and   $\varepsilon \; [\equiv - \dot{H}/H^2]$ being, 
respectively, the Hubble and slow-roll parameters, and $\dot{H} \equiv dH/dt = H'/a$. Then, we find that,
\bqn
\lb{eomb}
&&\frac{d^2 \mu_{k}(y)}{dy^2}+\left[\hat\omega_k^2(\eta)
 -\frac{\nu^2(\eta)-1/4}{y^2}\right]\mu_k(y)=0, ~~~~
\eqn
where $y\equiv -k\eta$ and
\bq
\lb{omegaB}
 \hat\omega_k^2(\eta) \equiv  \frac{\omega_k^2(\eta)}{k^2} = 1-b_1\epsilon_*^2 y^2 + b_2 \epsilon_*^4 y^4, 
 \eq
 with $\epsilon_* \equiv H/M_*, \;  b_1\equiv  (1+2\varepsilon)\hat{b}_1,\;
b_2\equiv (1+4 \varepsilon)  \hat{b}_2$,  and
\bqn
\lb{ab}
\frac{z''}{z} &\equiv& \frac{\nu^2(\eta)-1/4}{\eta^2}.
\eqn
To the first-order approximations of the slow-roll inflation, one can treat $\nu(\eta)$ and the parameters $b_1$ and $b_2$ as
constants   \cite{Brandenberger2013CQG}.

Obtaining  analytically approximate solutions of Eq.\ (\ref{eomb}) with proper initial conditions  is one of the crucial steps  in the 
understanding  of quantum effects on inflation and cosmological perturbations, and has been intensively investigated in the past 
decade \cite{Brandenberger2013CQG}. These studies  were carried out mainly by using the  Brandenberger-Martin method, in which 
the evolution of $\mu_k(\eta)$ is divided into several epochs, and in each of them the solution can be obtained either by  the WKB 
approximations when the adiabatic condition 
\bq
\lb{adiabaticC}
\left|\frac{3 \omega'^2_k}{4 \omega_k^4} - \frac{\omega''_k}{2 \omega_k^3}\right| \ll 1,
\eq
is satisfied, or by the linear combination of the exponentially decaying and growing modes, otherwise. Then, the individual solutions 
were matched together at the boundaries.  While this often yields reasonable analytical approximations, its validity in various physical 
situations has been questioned recently  \cite{JM09,ACD}, and shown that it is valid only when the comoving wavenumber $k \gg aH$. 
In addition, in this method the error bounds are not known, and can be obtained only by comparing it with the numerical (exact) evolution 
of the mode function in the case-by-case basis. However, with the arrival of the precision era of cosmological measurements, accurate 
calculations of cosmological variables  are highly demanded \cite{KDM}.

In this paper,   we propose another method --- {\em the uniform asymptotic approximation}, to construct accurate analytical solutions of the 
mode function $\mu_{k}(y)$, and present explicitly the  error  bounds   for the error terms associated with the approximations. It is precisely 
because of the understanding and control of the errors that such constructed approximate solutions describe the exact evolutions of the 
perturbations extremely well, even only to the first-order approximations. As an application of the approximate analytical solutions, 
we calculate the power spectra and indices of scalar and tensor perturbations in the slow-roll inflation, and find that the amplitudes 
of the power spectra get modified due to the quantum effects, while the power spectrum indices remain the same as in the linear 
case.  Specifically,   the paper is organized as follows: In Section II, using the  Liouville transformations we first write Eq.(\ref{eomb}) in a proper
form.  Then, applying  the uniform asymptotic approximations we construct approximate solutions and  the corresponding error bounds, and study 
them in details. Thanks to the understanding of the error bounds, the  Liouville transformations are chosen so that the errors are minimized, and
the analytical solutions describe the exact evolution of the perturbations extremely well, as one can see from Figs. \ref{fig1} - \ref{fig4}.
In Section III, the particular case  $|\xi_0|^2 \gg1 $ is considered. In Section IV, 
as a simple application of the approximate  analytical solutions constructed in Section III and IV,  we calculate the power
spectra and  indices of scalar and tensor perturbations, by paying particular attention on the  modifications of them due to the
quantum  effects. In Section V, we present our main conclusions. Five appendices  are also included, in which some useful
mathematical  expressions and detailed calculations of quantities involved in the context are given. 

It should be noted that a relevant consideration  was previously carried out by Habib {\em et al} \cite{uniformPRL} for the 
inflationary universe in GR, in which the dispersion relation is linear, given by Eq.(\ref{omega}) with $\hat b_i = 0$. Clearly, 
their treatment  is applicable only to the case where the function $g(\eta)$  defined in Eq.(\ref{function}) has only one single turning point 
(or a single root of the equation $g(\eta) = 0$). It cannot be applied to the more interesting cases with several turning points, in particular,  
to those where some turning points may be double, triple or even higher-multiple  roots. 
The method to be proposed in this paper  is  to treat all the above cases in a unified way, which is mathematically quite different from that of 
\cite{uniformPRL}, and reduces to it when $\hat b_i = 0$. Part of the results was reported in \cite{ZWCKS}.

\section{Liouville transformations and approximate analytical solutions}
\renewcommand{\theequation}{2.\arabic{equation}} \setcounter{equation}{0}

Following \cite{Olver1974,Olver1975,Nayfeh}, we first write  Eq.(\ref{eomb}) in the form
\bqn\lb{eom58}
\frac{d^2 \mu_k (y)}{dy^2}=\big[g(y)+q(y)\big]\mu_k(y),
\eqn
where
\bqn\lb{gplusq}
g(y)+q(y)\equiv \frac{\nu^2-1/4}{y^2} -1+b_1 \epsilon_*^2 y^2-b_2 \epsilon_*^4 y^4,
\eqn
and the functions $g(y)$ and $q(y)$ will be determined by the analysis of the error bounds given below, so that the associated errors will be minimized. 
Clearly,  $g(y)$ and $q(y)$ in general have two poles, one is at $y= 0^{+} $ and  the other is at $y= +\infty$. In addition,   the function $g(y)$ can vanish 
at various points, which are called  the turning points or zeros, according to  the terminology used in \cite{Olver1974,Olver1975,Nayfeh}. From the theory of the 
second-order differential equations, the asymptotic solutions of the mode function $\mu_{k}(y)$ depend on the behavior of functions $g(y)$ and $q(y)$ around the poles and turning 
points. In the following, we present a technique for determining the analytic approximate solutions around them in a unified way.

 To such goals, let us first introduce the Liouville transformations with two new variables $U(\xi)$ and $\xi$,  
\bqn
\lb{Olver trans}
U(\xi)&=& \chi^{1/4} \mu_k(y),\;\;\; \xi'^2 =  \frac{|g(y)|}{f^{(1)}(\xi)^2},
\eqn
where $ \chi \equiv \xi'^2,\; \xi'=d\xi/dy$, and 
\bqn
\lb{OlverTransB}
f(\xi)&=& \int^y \sqrt{|g(y)|} dy,\;\;\;  f^{(1)}(\xi)=\frac{df(\xi)}{d\xi}.
\eqn
Note that $\chi$ must be regular and not vanish in the intervals of interest. Consequently, $f(\xi)$ must be chosen so that 
$f^{(1)}(\xi)$ has zeros and singularities of the same type   as that of $g(y)$. As shown below, such requirements play an 
essential role in determining the approximate solutions.
 In terms of $U$ and $\xi$, Eq.(\ref{eom58}) takes the form
\bqn\lb{eomU}
\frac{d^2 U}{d\xi^2}&=&\left[\pm f^{(1)}(\xi)^2+\psi(\xi)\right]U,  
\eqn
where
\bqn\lb{psi}
\psi(\xi)=\frac{q(y)}{\chi}-\chi^{-3/4} \frac{d^2(\chi^{-1/4})}{dy^2},
\eqn
and the signs ``$\pm$" correspond to $g(y)>0$ and $g(y)<0$, respectively. Considering $\psi(\xi) =0$ as the first-order approximation,
one can choose $f^{(1)}(\xi)$ so that the first-order approximation can be as close
to the exact solutions as possible with the guidelines of the error functions constructed below, and then  solve it in terms of known functions.
Clearly, such a choice  sensitively  depends on the behaviors of the functions $g(y)$ and $q(y)$ near  the poles and turning points.

\subsection{Liouville-Green Solutions Near Two Poles}

From Eq.(\ref{gplusq}), we can see that, except exactly at  the  two poles, located at  $y= 0^{+}$ and $y= +\infty$, respectively,   the functions $g(y)$ and $q(y)$ 
are well-defined  in their neighborhoods. With this property, we can choose
\bqn\lb{fpole}
f^{(1)}(\xi)^2=\text{const}.
\eqn
Without loss  of the generality, we can always set  this constant to one. Then,  from Eq.(\ref{Olver trans}) we find 
\bqn
\xi=\int^{y} \sqrt{\pm g(y)} dy,
\eqn
here ``+" ( ``-") corresponds to the pole $y = 0^{+}$  ($y = +\infty$), and  the equation of motion (\ref{eomU})  takes the form
\bqn\lb{eompole}
\frac{d^2U}{d\xi^2}=\left[\pm 1+\psi(\xi)\right] U.
\eqn

Let us first consider the approximate solution near the pole $ y = 0^{+}$. As the first-order approximation, neglecting  the $\psi(\xi)$ term,  from  Eq. (\ref{eompole})
we find 
\bqn\lb{Uplus}
U^{+}=c_{+} e^{\xi} (1+\epsilon^{+}_1)+d_{+}e^{-\xi} (1+\epsilon^{+}_2),
\eqn
where $\epsilon^{+}_1$ and $\epsilon^{+}_2$ represent the errors of the approximate solution. Accordingly, the mode function $\mu_k(y)$ is given by  the 
Liouville-Green (LG) solution
\bqn\lb{LG1}
\mu^{+}_k(y)&=& \frac{c_{+}}{g(y)^{1/4}} e^{\int^y \sqrt{g(y)}dy} (1+\epsilon^{+}_1) \nb\\
&&\;\;\;+ \frac{d_{+}}{g(y)^{1/4}} e^{-\int^y \sqrt{g(y)}dy} (1+\epsilon^{+}_2).
\eqn

Similarly, near the pole $ y = +\infty$, Eq.(\ref{eompole}) has the solution
\bqn\lb{Uminus}
U^{-}= c_{-} e^{i \xi }(1+\epsilon_1^{-})+d_{-} e^{-i \xi} (1+\epsilon_2^{-}),
\eqn
where $\epsilon^{-}_1$ and $\epsilon^{-}_2$ represent the errors of the approximations, and the mode function $\mu^{-}_k(y)$ takes the form
\bqn\lb{LG2}
\mu^{-}_k(y)&=&\frac{c_{-}}{(-g(y))^{1/4}} e^{i \int^y \sqrt{-g(y)}dy} (1+\epsilon^{-}_1)\nb\\
&&+\frac{d_{-}}{(-g(y))^{1/4}} e^{-i \int^y \sqrt{-g(y)}dy} (1+\epsilon^{-}_2). ~~~~~
\eqn

\subsubsection{Error bounds of  the LG solutions}
Obviously,  the accuracy of the approximate solutions (\ref{Uplus}) and (\ref{Uminus}) depends on the magnitude of $\psi(\xi)$, which was neglected when solving Eq.(\ref{eompole})
to the first-order approximations. With the choice of Eq.(\ref{fpole}), we find
\bqn
\psi(\xi)=\frac{q(y)}{|g(y)|}-\frac{1}{|g(y)|^{3/4}} \frac{d^2}{dy^2}\left(\frac{1}{|g(y)|^{1/4}}\right).
\eqn
Roughly speaking, the resulting approximations are meaningful when $\psi(\xi)$ is much smaller than one, which shall be true, provided that
\bqn
\lb{conditionA}
 \left|\frac {q(y)}{g(y)} \right| \ll 1,\;\;\;
 \left||g|^{-3/4}\frac{d^2(|g^{-1/4}|)}{dy^2}\right| \ll 1.
\eqn
 These two conditions give rise to strong constraints on the choice of the  functions $g(y)$ and $q(y)$. To see these, 
 we first consider the error $\epsilon_1^{+}$. Substituting the first branch of $U^{+} \sim e^{\xi} (1+\epsilon_{1}^{+})$ into Eq.(\ref{eompole}), 
we  obtain a second-order differential equation for $\epsilon_1^{+}$,
\bqn\lb{error1plus}
\frac{d^2\epsilon_1^{+}}{d\xi^2}+2 \frac{d \epsilon_1^{+}}{d\xi}=\psi(\xi) (1+\epsilon_1^{+}).
\eqn
By using the variations of parameters we find that the error term $\epsilon_1^{+}$ has the solution
\bqn
\epsilon_{1}^{+}=\frac{1}{2}\int_{0}^{\xi} \left(1-e^{2(v-\xi)}\right)\psi(v) (1+\epsilon_1^{+})dv,
\eqn
where $\xi \in (0,a_1)$, $v\in (0,\xi]$, and $y\in (0^{+},\hat{a}_1)$ with $a_1,\;\hat{a}_1$ are the upper bounds of variable $\xi$ and $y$, 
respectively, and  $\xi(0^{+})=0,\; \xi(\hat{a}_1)=a_1$.
Comparing Eq.(\ref{error1plus}) with Eq.(\ref{error}), we obtain
\bqn
{\cal{K}}(\xi,v)=\frac{1}{2} \left(1-e^{2(v-\xi)}\right),\;\;\;J(v)=1,\nb\\
\phi(v)=\psi_0(v)=\psi(v),\;\;\psi_1(v)=0.
\eqn
Considering  $0< v\leq \xi$,  we find 
\bqn
\left|{\cal{K}}(\xi,v)\right| \leq \frac{1}{2},\;\;\left|\frac{\partial {\cal{K}}(\xi,v)}{\partial \xi}\right|\leq 1.
\eqn
Hence, we obtain,  
\bqn
P_0(\xi)=\frac{1}{2},\;\;\;Q(v)=1,\;\;P_1(\xi)=1,
\eqn
where $P_i$ and $Q$ are all defined in Appendix A. Thus,  applying the theorems presented in Appendix A, the error bounds for $\epsilon_1^{+}$ read
\bqn
|\epsilon_1^{+}|,\;\;\;\frac{|d\epsilon_1^{+}/dy|}{2 |g|^{1/2}} \leq \exp{\left(\frac{1}{2} \mathscr{V}_{0,y}(F)\right)}-1,
\eqn
where $F(y)$ is the error control function, defined as
\bqn\lb{error infty}
F(y)&=&\int |\psi(v)|dv\nb\\
&=&\int \left[\frac{1}{|g|^{1/4}}\frac{d^2}{dy^2}\left(\frac{1}{|g|^{1/4}}\right)-\frac{q}{|g|^{1/2}}\right] dy, ~~~~~
\eqn
and $\mathscr{V}_{x_1,x_2}(F)$ is defined as
\bqn
\mathscr{V}_{x_1,x_2}(F) \equiv \int^{x_2}_{x_1} \left|\frac{dF(y)}{dy}\right|dy,
\eqn
representing the total variation of the function $F(y)$. 

Similarly, it can be shown that the error bounds for the error $\epsilon_2^{+}$ are given by
\bqn
|\epsilon_2^{+}|, \frac{\left|d\epsilon_2^{+}/dy\right|}{2|g|^{1/2}} \;\leq \exp{\left(\frac{1}{2}\mathscr{V}_{y,\hat{a}_1}(F)\right)}-1.
\eqn

Repeating the above analysis for the error bounds near the pole $y = +\infty$, we find that   the error bounds for $\epsilon_{1,2}^{-}$ take the forms,
\bqn
|\epsilon_{1}^{-}|,\;\frac{|d\epsilon_1^{-}/dy|}{|g|^{1/2}} \leq \exp{\left(\mathscr{V}_{y,+\infty}(F)\right)}-1,\nb\\
|\epsilon_{2}^{-}|,\;\frac{|d\epsilon_2^{-}/dy|}{|g|^{1/2}} \leq \exp{\left(\mathscr{V}_{\hat{a}_2,y}(F)\right)}-1,
\eqn
where $\hat{a}_2$ is the lower bound of $y$, i.e., $y\in (\hat{a}_2,+\infty)$.

\subsubsection{Choice of Functions $g(y)$ and $q(y)$ }

At the pole $ y = 0^{+}$,  as pointed out in \cite{Olver1974}, the LG approximations are valid  only when the function $g(y)$ has a pole of order $m\geq 2$. 
Assuming that $q(y)$ also has a pole at this point, but of order $n$, then we can expand them in the form, 
\bqn
\lb{expand function}
g(y)=\frac{1}{y^m}\sum_{s=0}^\infty g_s y^{s},\;q(y)=\frac{1}{y^n} \sum_{s=0}^\infty q_s y^s.
\eqn
When $m>2$, from Eq.(\ref{gplusq}) we can see that we must have $n=m$, and consequently $q_0=-g_0$. Thus,  near the pole $y = 0^+$, 
the terms $g_0/y^m$ and $q_0/y^m$ dominate, and the   condition $|q(y)|\ll |g(y)|$ [cf. Eq.(\ref{conditionA})] is violated. Therefore, we must set $m = 2$, for which 
the condition  $|q(y)|\ll |g(y)|$ requires   $n\leq 2$. Then,   substituting  Eq.(\ref{expand function}) into the integrand of the error control function $F(y)$, we obtain
\bqn
&&\frac{1}{|g|^{1/4}} \frac{d^2}{dy^2} \left(\frac{1}{|g|^{1/4}}\right)-\frac{q}{|g|^{1/2}} \simeq \frac{1}{y} \left[-\frac{1}{4 |g_0|^{1/2}} \right.\nb\\
&&  ~~~~~~~~~   \left. +\frac{49}{8}\frac{|g_1|}{|g_0|^{3/2}} y+{\cal{O}}(y^2) \right] -\frac{1}{y^{n-1}} \left[\frac{q_0}{|g_0|^{1/2}}\right.\nb\\
&& ~~~~~~~~~ \left. +\left(q_1|g_0|^{1/2}-\frac{|g_1|q_0}{2|g_0|^{3/2}}\right)y +{\cal{O}}(y^2)\right].~~~~
\eqn
Clearly, the convergence of the error control function $F(y)$ leads to  
\bqn\lb{c1}
n=2, \;\;\;\;\;\;q_0=-\frac{1}{4}.
\eqn
On the other hand, the terms $q_1/y$ and $g_1/y$ violate the condition   $|q(y)|\ll |g(y)|$, unless  they vanish identically, 
\bqn\lb{c2}
q_1=0=-g_1.
\eqn

Let us now turn to the pole $y = +\infty$. Assuming that $g(y)$ and $q(y)$ have a pole of order $\bar{m}$  and $\bar{n}$, respectively, we find that they can be 
expanded into the forms,  
\bqn
\lb{function expand}
g(y)=y^{\bar{m}} \sum_{s=0}^\infty \bar{g}_s y^{-s},\; q(y)=y^{\bar{n}} \sum_{s=0}^\infty \bar{q}_s y^{-s},
\eqn
where $\bar{g}_s$ and $\bar{q}_s$ are other sets of constants.   Substituting the above expansions into the integrand of the error control function $F(y)$, we find that
\bqn
&&\frac{1}{|g|^{1/4}} \frac{d^2}{dy^2} \left(\frac{1}{|g|^{1/4}}\right)-\frac{q}{|g|^{1/2}} = y^{-2-\bar{m}/2} \sum_{s=0}^{+\infty} c^{(1)}_s y^{-s}\nb\\
&&~~~~~~~~~~~~ +y^{\bar{n}-\bar{m}/2} \sum_{s=0}^{+\infty} c_s^{(2)} y^{-s},
\eqn
where the coefficients $c_s^{(1)}$ and $c_s^{(2)}$ are functions of $\bar{q}_s$ and $\bar{g}_s$. Then,  the convergence of  $F(y)$ requires
\bqn
\lb{ConditionB}
\bar{m}>-2, \;\;\;\bar{n}<\frac{\bar{m}}{2}-1.\nb
\eqn
When $\bar{m}>4$ or $\bar{m}<4$, from Eq.(\ref{gplusq}) one finds  that  $\bar{n}$ must be taken either $\bar{n}=\bar{m}$ or $\bar{n}=4$. But, these
shall  violate the above conditions just mentioned.  Therefore, we must choose $\bar{m} = 4$, for which $ \bar{n}$ must be less than 1, that is, 
\bqn
\lb{c3}
\bar{m}=4,\;\;\; \bar{n}<1.
\eqn

Combining the conditions given by Eqs.(\ref{c1}), (\ref{c2}) and (\ref{c3}), we find that  the function  $q(y)$ must take the form, 
$q(y) = -{1}/({4y^2}) +q_2$ with $q_2+g_2=-1$ and $|q_2|\ll |g_2|$ \footnote{In principle, it is not necessary to choose $g(y)$ and $q(y)$ in the same forms 
near both of the two poles. For example, one can take a choice of $g(y)$ and $q(y)$ which only satisfies Conditions (\ref{c1}) and (\ref{c2}) near the pole $ y=0^+$, 
while taking another choice of $g(y)$ and $q(y)$ that  only satisfies Condition (\ref{c3}) near the pole $ y = +\infty$. In the intermediate region between the
 two poles, the differential equation (\ref{eom58}) will has turning points that depend on the explicit form of $g(y)$. As  shown below,  we will have 
 approximate solutions near these turning points that are different from the LG solutions, but must  go over to them   near these two poles.
For such approximate solutions, the convergence of the errors requires that  $g(y)$ and $q(y)$ must satisfy all the conditions of Eqs.(\ref{c1}), (\ref{c2}), and (\ref{c3}). 
As a result, the function  $q(y)$ is necessarily to take the form $q(y) = -{1}/({4y^2}) +q_2$. Then, Eq.(\ref{function}) follows.}. 
Without loss of the generality, we can always set $q_2=0$, so finally we obtain  
\bqn
\lb{function}
g(y)&=&\frac{\nu^2}{y^2}- 1 +b_1 \epsilon_*^2 y^2 -b_2 \epsilon_*^4 y^4,\nb\\
q(y)&=&-\frac{1}{4y^2}.
\eqn
Clearly, depending on values of $\nu$ and $b_i$, the function $g(y)$ has different behavior. Fig. \ref{fig0}
shows these possibilities.

\begin{figure}[t]
\centering
	{\includegraphics[width=75mm]{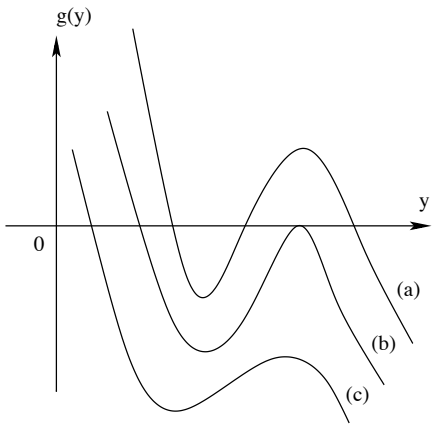}}
\caption{The function $g(y)$ defined by Eq.(\ref{function}).  
(a)  Three  different  single real  roots of the equation $g(y) = 0$. 
(b) One single and one double real roots. 
(c)  One single   real root.}
\lb{fig0}
\end{figure}

\subsection{Classification of Turning Points }

The LG approximate solutions only apply to the region that $g(y)$ does not vanish. Once $g(y)$ is zero, both  $\psi(\xi)$ and $\mu_k(y)$ 
becomes divergent near these points, and the LG approximations  fail to hold. In order to get the asymptotic solutions around these points, 
one must take  different choices of $f^{(1)}(\xi)^2$ in the Liouville transformations (\ref{Olver trans}). However, such choices depend on the 
natures of the turning points. Therefore, we first investigate the natures of them.
In general, the equation $g(y)=0$ is a cubic equation, and  can be always cast in the form,
\bqn
\lb{cubic}
b_2 x^6 -b_1 x^4+x^2-\nu^2 \epsilon_*^2=0,
\eqn
where $x=\epsilon_* y >0$. Setting  ${\cal{Y}}\equiv 3 b_2/ b_1^2$, the three roots of Eq.(\ref{cubic}) are given by
\bqn
\lb{roots}
x_0&=& \left\{\frac{b_1}{3b_2} \left[1-2\sqrt{1- {\cal{Y}}} \cos{\left(\frac{\theta}{3}\right)}\right]\right\}^{1/2},\nb\\
x_1&=& \left\{\frac{b_1}{3b_2} \left[1-2\sqrt{1- {\cal{Y}}} \cos{\left(\frac{\theta+2\pi}{3}\right)}\right]\right\}^{1/2},\nb\\
x_2&=& \left\{\frac{b_1}{3b_2} \left[1-2\sqrt{1- {\cal{Y}}} \cos{\left(\frac{\theta+4\pi}{3}\right)}\right]\right\}^{1/2}, ~~~~~~~
\eqn
where
\bqn
\lb{rootA}
\cos{\theta} \equiv - \left(1-\frac{3}{2} {\cal{Y}}+\frac{3}{2} b_1 {\cal{Y}}^2 \nu^2 \epsilon_*^2 \right)(1-{\cal{Y}})^{-3/2}. ~~
\eqn

Depending on the signs of $\Delta$, the natures of the three roots  are different, where $\Delta$ is defined as
\bqn
\Delta\equiv ({\cal{Y}}-1)^3+\left(1-\frac{3}{2} {\cal{Y}}+\frac{3}{2} b_1 {\cal{Y}}^2 \nu^2 \epsilon_*^2\right)^2.
\eqn
In particular, when $\Delta <0$,  the three roots $x_i$ are real and different [Fig. \ref{fig0}(a)]. 
When $\Delta = 0$, it  has one single real root and one double real root [Fig. \ref{fig0}(b)].
When $\Delta > 0$, it  has one single real root and two complex    roots [Fig. \ref{fig0}(c)].

In terms of $y$,   the three roots will be denoted by $y_i\; (i = 0, 1, 2)$. Without lose of the generality, we further assume $y_0<y_1\leq y_2$ in the case  
$\Delta <0$;  $y_0 < y_1$ and $y_2= y_1$ in the case $\Delta = 0$; and $ y_0$ is real and  $y_{1,2}$ are  complex with $y_1 = y_2^*$ in the case
$\Delta > 0$. Note that with such identifications, in general we do not have the relations $y_i = x_i/\epsilon_*$, where $x_i$ are the roots given by
Eq.(\ref{roots}).   Then, in all the three cases, we have $y_0 \sim {\cal{O}}(1)$, while the magnitude of roots $y_2$ and $y_1$ (for complex roots, $\text{Re}(y_1)$ and 
$\text{Re}(y_2)$) are dependent on the values of $\epsilon_*$. When $\epsilon_*\ll 1$, we have $y_{1,2} \gg 1$ ($\text{Re}(y_{1,2}) \gg 1$ for complex roots). 
However, the difference  between $y_1$ and $y_2$ generally depends on the choice of the free parameters appearing in the dispersion relations. 
 
To process further, let us specify the three conditions \cite{Olver1974,Nayfeh,Olver1975,Zhang1991}:
\begin{itemize}
\item[(a)]  $|q(y)| < |g(y)|$ holds everywhere, except in the neighborhoods of the turning points $y_i$; 
\item[(b)] $|q(y)| < \left|\frac{g(y)}{(y-y_i)}\right|$ holds in the neighborhoods of the turning points $y_i$; 
\item[(c)] $|q(y)| <  \left|\frac{g(y)}{(y-y_1)(y-y_2)}\right|$ holds in the neighborhoods of the turning points $y_1$ and $y_2$.
\end{itemize}
As we shall show below, these three conditions play essential roles in determining the approximate solutions around the turning points.

\subsection{Asymptotic Solution Around $y = y_0$}

In the neighborhood of $y_0$, Conditions (a) and (b) are satisfied, and $y_0$ is a single turning point. Then,
 one can introduce a monotone increasing or decreasing function $\xi$ via the relations, 
\bqn\lb{f1}
f^{(1)}(\xi)^2=\pm\xi,
\eqn
where $\xi(y_0)=0$. Without loss of the generality, we can choose $\xi$ to have the same sign as $g(y)$, and thus $\xi$ is a 
monotone decreasing function around   $y = y_0$. Combining Eqs.(\ref{Olver trans}) and (\ref{f1}), we find
\bqn
\lb{xiy0}
\xi=\begin{cases} 
                   - \left(\frac{3}{2} \int^y_{y_0} \sqrt{-g(y)} dy\right)^{\frac{2}{3}},& y\geq y_0,\\ 
                   \left(-\frac{3}{2} \int^y_{y_0} \sqrt{g(y)} dy\right)^{\frac{2}{3}},& y \leq y_0,
\end{cases} 
\eqn
and that Eq(\ref{eomU}) now reads, 
\bqn
\lb{eomy0}
\frac{d^2U}{d\xi^2}=\Big(\xi+\psi(\xi)\Big) U.
\eqn
Neglecting the $\psi(\xi)$ term as the first-order approximation, we find that the above equation has the approximate analytical solution
\bqn
U(\xi)=\alpha_0 \Big(\text{Ai}(\xi) +\epsilon_3\Big)+\beta_0 \Big(\text{Bi}(\xi)+\epsilon_4 \Big),
\eqn
where $\text{Ai}(\xi)$ and $\text{Bi}(\xi)$ are the Airy functions, $\alpha_0$ and $\beta_0$ are integration constants, 
and $\epsilon_3$ and $\epsilon_4$ denote the errors of the approximations.    Then,  the mode function is given by,  
\bqn\lb{airy solution1}
\mu_k(y)&=& \alpha_0  \left(\frac{\xi}{g(y)}\right)^{1/4}\Big(\text{Ai}(\xi) +\epsilon_3\Big)\nb\\
&&+\beta_0 \left(\frac{\xi}{g(y)}\right)^{1/4}\Big( \text{Bi}(\xi) +\epsilon_4\Big).
\eqn

To estimate the corresponding errors, let us first  consider $\epsilon_3$. Inserting   
$U(\xi)\sim \text{Ai}(\xi)+\epsilon_3$ into Eq.(\ref{eomy0}), we have 
\bqn
\frac{d^2 \epsilon_3}{d\xi^2} = \xi \epsilon_3+\psi(\xi)\Big( \text{Ai}(\xi) +\epsilon_3\Big).
\eqn
Treating the second term in the right-hand side of the above equation as corrections,  and then using the method of variation of parameters, 
we obtain
\bqn
\epsilon_3(\xi)=\int_{\xi}^{a_3} {\cal{K}}(\xi,v) |v|^{-1/2}\psi(v) \Big(\text{Ai}(v)+\epsilon_1\Big), \lb{error1}
\eqn
where $a_3$ is the upper bound of $\xi$ and the corresponding upper bound of $y$ is $\hat{a}_3$, i.e., $\xi(\hat{a}_3)=a_3$, and 
\bqn
{\cal{K}}(\xi,v)=\pi |v|^{1/2} \Big(\text{Bi}(\xi) \text{Ai}(v)-\text{Ai}(\xi) \text{Bi}(v)\Big).
\eqn
Bounds of ${\cal{K}}(\xi,v)$ and $\partial {\cal{K}}(\xi,v)/\partial \xi$ are expressible in terms of the auxiliary functions of the Airy 
functions (See Appendix B for details). Thus, we find 
\bqn
|{\cal{K}}(\xi,v)|&=&\pi \sqrt{|v|} M(\xi)M(v)\Bigg|\frac{E(\xi)}{E(v)}\cos{\theta(\xi)}\sin{\theta(v)}\nb\\
&&\;\;\;\;\;-\frac{E(v)}{E(\xi)}\cos{\theta(v)}\sin{\theta(\xi)}\Bigg|\nb\\
&\leq& \frac{\pi \sqrt{|v|}M(\xi)M(v)E(v)}{E(\xi)},
\eqn
for $\xi\leq v\leq a_3$, where $E(x)$ is a non-decreasing function of $x$. We also have
\bqn
\left|\frac{\partial {\cal{K}}(\xi,v)}{\partial \xi}\right| \leq \pi \sqrt{|v|}N(\xi) M(v)E(v)E^{-1}(\xi).
\eqn 
Then,  from the above we obtain
\bqn
Q(v)&=& \pi |v|^{1/2}E(v) M(v),\;\;\nb\\
P_0(\xi)&=&E^{-1}(\xi)M(\xi),\nb\\
P_1(\xi)&=&E^{-1}(\xi)N(\xi).
\eqn
Comparing Eq.(\ref{error1}) with Eq.(\ref{error}), we find 
\bqn
\phi(v)=\psi_0(v)=|v|^{-1/2} \psi(v), \nb\\
\psi_1(v)=0,\;\;J(v)=\text{Ai}(v).
\eqn
From Appendix A, the quantities $\kappa,\;\kappa_0$ now can be calculated and are given by
\bqn
\kappa&=&\text{sup}\{\pi |v|^{1/2}E(\xi) M(\xi) |\text{Ai}(\xi)|\},\nb\\
\kappa_0&=&\text{sup}\{\pi |v|^{1/2}M^2(\xi)\}.
\eqn
Numerically, we find that these  yield $\kappa=1,\;\;\kappa_0=1.03952$.

With the above results, the error bounds for $\epsilon_3$ read
\bqn\lb{error3}
\frac{|\epsilon_3|}{M(\xi)},\;\frac{|\partial \epsilon_3/\partial \xi|}{ N(\xi)} \leq \frac{ E^{-1}(\xi)}{\lambda} \Big\{\exp{\Big(\lambda \mathscr{V}_{\xi,a_3}(H)\Big)}-1\Big\},\nb\\
\eqn
where $\lambda\equiv \kappa_0$ and $H(\xi)$ is the corresponding error control function, give by
\bqn
\lb{FunctionH}
H(\xi)=\int^{a_3}_\xi |v|^{-1/2} \psi(v) dv.
\eqn
Similarly,  the error bound for $\epsilon_4$ can be expressed as, 
\bqn\lb{error4}
\frac{|\epsilon_4|}{M(\xi)},\;\frac{|\partial \epsilon_4/\partial \xi|}{ N(\xi)} \leq \frac{ E(\xi)}{\lambda} \Big\{\exp{\Big(\lambda \mathscr{V}_{a_4,\xi}(H)\Big)}-1\Big\},\nb\\
\eqn
where $a_4$ is the lower bound of the variable $\xi$, and the corresponding upper bound for $y$ is $\hat{a}_4$, i.e., $\xi(\hat{a}_4)=a_4$. Note that for $\epsilon_4$, we have
\bqn
Q(v)& =&\pi |v|^{1/2} E^{-1}(v) M(v),\nb\\
P_0(\xi) &=& E(\xi) M(\xi),\;\;\; P_1(\xi) = E(\xi) N(\xi),\nb\\
J(v)&=& \text{Bi}(v),\nb\\
\kappa&=&\text{sup}\{\pi |v|^{1/2} E^{-1}(\xi) M(\xi) |\text{Bi}(\xi)|\},
\eqn
which are different from those for $\epsilon_3$. Numerical calculations also yield $\kappa=1$.

\subsection{Asymptotic Solution Near   $y_1$ and $y_2$}

Depending on $\Delta < 0, \; \Delta = 0$ and $ \Delta > 0$, the roots $y_1$ and $y_2$ have  different properties, as shown above. Although these three 
different cases can be teated in a unfired way, let us first consider them,  separately. 

\subsubsection{ When $y_{1}$ and $y_{2}$ Are Real}

In this case, near the turning points $y_1$ and $y_2$, Conditions (a) and (c) are always satisfied. When Condition (b) is also satisfied, 
one can treat $y_1$ and $y_2$ as simple turning points, and similar to that near the turning point $y_0$, one can get the asymptotic solutions near these points. However, 
when $|y_2-y_1|\ll 1$, Condition (b) is not satisfied, and then the method used for the turning point $y_0$ is no longer valid. 
Following Olver\cite{Olver1975}, we shall adopt a method to treat  these cases together. The crucial point is to choose $f^{(1)}(\xi)^2$ in the Liouville transformations (\ref{Olver trans})
as
\bqn
\lb{xi0A}
f^{(1)}(\xi)^2=|\xi^2-\xi_0^2|,
\eqn
where $\xi$ is an increasing function of $y$,  and with the choices $\xi(y_1)=-\xi_0$ and $\xi(y_2)=\xi_0$. When $\xi_0=0$, it corresponds to 
the case  $y_1=y_2$. Then, we find that in  the region  $y\geq y_2 $ we have $g(y)<0$ and $\xi\geq \xi_0$, and 
\bqn
\lb{xi2}
&& \int_{y_2}^{y} \sqrt{-g(y)} dy=\frac{1}{2} \xi \sqrt{\xi^2-\xi_0^2}-\frac{1}{2} \xi_0^2 \text{arcosh}\left(\frac{\xi}{\xi_0}\right),\nb\\
&& ~~~~~~~~~~~~~~~~~~~~~~~~~~~~~~~~~~~~~~~~~~~~~~~ (y \geq y_2).
\eqn
Similarly, in the region  $y \leq y_1$, we have $g(y)>0$ and $\xi\leq -\xi_0$, while in the region $y \in (y_1,y_2)$, we have $g(y)>0$ and $-\xi_0 \leq \xi\leq \xi_0$.
Thus, we find that
\bqn
\lb{xi1}
 \int_{y_1}^{y} \sqrt{g(y)} dy&=& \frac{1}{2} \xi_0^2 \text{arcosh}\left(-\frac{\xi}{\xi_0}\right)  \nb\\
&& + \frac{1}{2}\xi \times
\begin{cases}
 \sqrt{\xi^2 - \xi_0^2 }, &  y \leq y_1,\\
\sqrt{\xi_0^2 -\xi^2},   & y \in (y_1,y_2),
\end{cases} ~~~~~~~
\eqn
 where  $\xi_0$ is given by
\bqn\lb{xi0}
\xi_0^2=\frac{2}{\pi} \int_{y_1}^{y_2} \sqrt{g(y)} dy.
\eqn
Then,  Eq.(\ref{eomU}) reduces to
\bqn
\lb{eomy1y2}
\frac{d^2U}{d\xi^2}=\left[\left(\xi_0^2-\xi^2\right)+\psi(\xi)\right]U.
\eqn
Neglecting the $\psi(\xi)$ term, we find that  the approximate solutions   can be expressed in terms of the parabolic cylinder functions $W(\frac{1}{2}\xi_0^2,\pm \sqrt{2} \xi)$,
and are given by
\bqn
U(\xi)&=& \alpha_1 \Bigg\{W\left(\frac{1}{2}\xi_0^2, \sqrt{2}\xi \right)+\epsilon_5\Bigg\}\nb\\
&&+\beta_1 \Bigg\{W\left(\frac{1}{2}\xi_0^2, -\sqrt{2}\xi \right)+\epsilon_6\Bigg\},
\eqn
for which we have
\bqn\lb{solutionW}
\mu_k(y)&=&\alpha_1 \left(\frac{\xi^2-\xi_0^2}{-g(y)}\right)^{1/4} W\left(\frac{1}{2}\xi_0^2, \sqrt{2}\xi \right)\nb\\
&&+\beta_1 \left(\frac{\xi^2-\xi_0^2}{-g(y)}\right)^{1/4} W\left(\frac{1}{2}\xi_0^2, -\sqrt{2}\xi \right),\nb\\
\eqn
where $\epsilon_5$ and $\epsilon_6$ are  the errors of the corresponding  approximates.

\subsubsection{When $y_1$ and $y_2$ Are Complex}

In this case, $y_1$ and $y_2$ are complex conjugate, $y_1 = y_2^*$. To have  a unified treatment,  here we still choose $f^{(1)}(\xi)^2$ in 
the Liouville transformations (\ref{Olver trans}) as \cite{Olver1975}
\bqn
\lb{xi0B}
f^{(1)}(\xi)^2=\xi^2-\xi_0^2,
\eqn
where $\xi$ is still an increasing variable and is chosen so that   $\xi(y_1)=- \xi_0$ and $\xi(y_2)=\xi_0$, respectively, but now with $\xi^2_0 < 0$, 
that is $\xi_0$ becomes imaginary.  When $\xi_0=0$, it reduces to the case   $y_1 = y_2$. Since now there are no real turning points, the function  $f^{(1)}(\xi)^2$ 
does not need to have any real zeros or singularities in the neighborhoods of $y=\text{Re}(y_1)$. Then, we can choose   
\bqn\lb{xi02}
\xi_0^2=- \frac{2}{\pi} \left|\int_{y_1}^{y_2} \sqrt{-g(y)}dy\right|.
\eqn
Note that in above integral the path lies along the imaginary axis, and the branch of $\sqrt{-g(y)}$ is real and positive. 
Since now $\xi_0$ is purely imaginary,  without loss of the generality, we set $\text{Im}(\xi_0)\geq0$.
Note that we also have  $g(y)=g(-y)$  for both real and purely imaginary values of $y$, we find 
\bqn
\int^{y} \sqrt{-g(y)}dy=\int^\xi \sqrt{\xi^2-\xi_0^2}d\xi.
\eqn
Then,  $\xi$ can be obtained from the integral,
\bqn
&& \int_{\text{Re}(y_1)}^y \sqrt{-g(y)}dy = \frac{1}{2}\xi \sqrt{\xi^2-\xi_0^2}\nb\\
&& ~~~~~~~~~~~~~~~~~~~~~~~~ - \frac{1}{2}\xi_0^2 \ln{\left(\frac{\xi+\sqrt{\xi^2-\xi_0^2}}{|\xi_0|}\right)},~~~~~~~
\eqn
and now the integral path lies along the real axis. With such chosen $\xi$, it can be shown that Eq.(\ref{eomU})  takes the same form as that of Eq.(\ref{eomy1y2}).
As a result,   the approximate solution is also  given by Eq.(\ref{solutionW}), but now $\xi_0$ is given by Eq.(\ref{xi02}).

The above shows clearly that the three cases ($\Delta > 0, \; \Delta= 0 , \; \Delta < 0$) can be treated in a unified way, although initially we considered them,  separately. 

\subsubsection{Error Bounds}

Since the approximate solutions are given by Eq.(\ref{solutionW}) in all the three  cases considered above, we do not need to distinguish them, when we consider 
the corresponding  errors $\epsilon_5$ and $\epsilon_6$. In particular, for $U$ given by
\bqn
U = W\left(\frac{1}{2}\xi_0^2,\sqrt{2}\xi\right) +\epsilon_5(\xi), 
\eqn
from Eq.(\ref{eomy1y2}), we find that  
\bqn
\frac{d^2 \epsilon_5}{d\xi^2} &=& -\left(\xi^2-\xi_0^2\right) \epsilon_5\nb\\
&&+\psi(\xi)\Big\{W \left(\frac{1}{2}\xi_0^2,\sqrt{2}\xi\right) +\epsilon_5\Big\}.
\eqn
Considering  the second term  in the right-hand side of the above equation as corrections, we find that    the error can be written in the form, 
\bqn\lb{error2}
\epsilon_5(\xi)= \int^{a_5}_\xi {\cal{K}}(\xi,v) \frac{\psi(v)}{v} \Big\{ W\left(\frac{1}{2}\xi_0^2,\sqrt{2}v\right)+\epsilon_5\Big\}dv,\nb\\
\eqn
where $a_5$ is the upper bound of $\xi$ and
\bqn
{\cal{K}}(\xi,v)&=& v  \Big\{W\left(\frac{1}{2}\xi_0^2,\sqrt{2}\xi\right) W\left(\frac{1}{2}\xi_0^2,-\sqrt{2}v\right)\nb\\
&&－W\left(\frac{1}{2}\xi_0^2,\sqrt{2}v\right) W\left(\frac{1}{2}\xi_0^2,-\sqrt{2}\xi\right)\Big\}. ~~~~
\eqn
Then,  in terms of the weight function $E\left(\frac{1}{2} \xi_0^2,\sqrt{2}\xi\right)$, the modulus function $M \left(\frac{1}{2}\xi_0^2,
\sqrt{2}\xi\right)$ for the function $W\left(\frac{1}{2}\xi_0^2,\sqrt{2}\xi\right)$ 
(See Appendix C for definitions of these auxiliary functions), we find that 
\bqn
|K(\xi,v)|&=&|v| M \left(\frac{1}{2}\xi_0^2,\sqrt{2}\xi\right) M \left(\frac{1}{2}\xi_0^2,\sqrt{2}v\right) \nb\\
&& \times \Bigg|\frac{E\left(\frac{1}{2} \xi_0^2,\sqrt{2}\xi\right)}{ E\left(\frac{1}{2} \xi_0^2,\sqrt{2}v\right)} \sin{\theta_1} \cos{\theta_2}\nb\\
&&\;\;-\frac{E\left(\frac{1}{2} \xi_0^2,\sqrt{2}v\right)}{ E\left(\frac{1}{2} \xi_0^2,\sqrt{2}\xi\right)} \sin{\theta_2} \cos{\theta_1}\Bigg|,\nb\\
\eqn
for $v\leq \xi\leq 0$, where $\theta_1\equiv \theta\left(\frac{1}{2} \xi_0^2,\sqrt{2} \xi \right)$, $\theta_2\equiv \theta\left(\frac{1}{2} \xi_0^2,\sqrt{2} v \right)$.
When the function $E\left(\frac{1}{2} \xi_0^2,\sqrt{2}\xi\right)$ is no-decreasing for $\xi\geq 0$, we also have
\bqn
|K(\xi,v)| &\leq &|v| M \left(\frac{1}{2}\xi_0^2,\sqrt{2}\xi\right) M \left(\frac{1}{2}\xi_0^2,\sqrt{2}v\right) \nb\\ 
&& \times  \frac{E\left(\frac{1}{2} \xi_0^2,\sqrt{2}v\right)}{ E\left(\frac{1}{2} \xi_0^2,\sqrt{2}\xi\right)}.
\eqn
Similarly, it can be shown that 
\bqn
\left|\frac{\partial K(\xi,v)}{\partial \xi}\right| &\leq& \sqrt{2} |v| N \left(\frac{1}{2}\xi_0^2,\sqrt{2}\xi\right) M \left(\frac{1}{2}\xi_0^2,\sqrt{2}v\right) \nb\\ 
&& \times  \frac{E\left(\frac{1}{2} \xi_0^2,\sqrt{2}v\right)}{ E\left(\frac{1}{2} \xi_0^2,\sqrt{2}\xi\right)},
\eqn
where $N\left(\frac{1}{2}\xi_0^2,\sqrt{2}\xi\right)$ is the modulus function of $W'\left(\frac{1}{2}\xi_0^2,\sqrt{2}\xi\right)$ and defined in Appendix C. Then, we
find that  
\bqn
Q(v)&=& |v| M \left(\frac{1}{2}\xi_0^2,\sqrt{2}v\right) E \left(\frac{1}{2}\xi_0^2,\sqrt{2}v\right),\nb\\
P_0(\xi) &=& \frac{M \left(\frac{1}{2}\xi_0^2,\sqrt{2}\xi\right)}{E \left(\frac{1}{2}\xi_0^2,\sqrt{2}\xi\right)},\nb\\
P_1(\xi)&=& \frac{N \left(\frac{1}{2}\xi_0^2,\sqrt{2}\xi\right)}{E \left(\frac{1}{2}\xi_0^2,\sqrt{2}\xi\right)}.
\eqn 
Comparing Eq.(\ref{error2}) with Eq.(\ref{error}), we obtain
\bqn
\phi(v)&=&\psi_0(v)=|v|^{-1} \psi(v), \nb\\
\psi_1(v)&=&0,\;\;J(v)=W\left(\frac{1}{2}\xi_0^2,\sqrt{2}v\right).
\eqn
With the above expressions, $\kappa$ and $\kappa_0$ are given by
\bqn
\kappa&=&\text{sup}\Bigg\{|\xi| M\left(\frac{1}{2}\xi_0^2,\sqrt{2}\xi\right) E\left(\frac{1}{2}\xi_0^2,\sqrt{2}\xi\right)\nb\\
&&\;\;\;\;\;\;\;\;\;\;\times \left|W\left(\frac{1}{2}\xi_0^2,\sqrt{2}\xi\right)\right|\Bigg\},\nb\\
\kappa_0&=&\text{sup}\Bigg\{|\xi| M^2\left(\frac{1}{2}\xi_0^2,\sqrt{2}\xi\right)\Bigg\}.
\eqn
Therefore,  for $\xi\leq 0$ the error bounds for $\epsilon_5$ can be expressed as
\bqn
&&\frac{|\epsilon_5|}{M\left(\frac{1}{2}\xi_0^2,\sqrt{2}\xi\right)},\;\frac{|\partial \epsilon_5/\partial \xi|}{\sqrt{2} N\left(\frac{1}{2}\xi_0^2,\sqrt{2}\xi\right)}\nb\\
&&\;\;\;\leq \frac{\kappa}{\lambda E\left(\frac{1}{2}\xi_0^2,\sqrt{2}\xi\right)} \Bigg\{\exp{\Big(\lambda \mathscr{V}_{\xi,a_5}(H)\Big)}-1\Bigg\}. ~~~~~~
\eqn

Similarly, it can be shown that the error bound for  $\epsilon_6$ are given by,
\bqn
&&\frac{|\epsilon_6|}{M\left(\frac{1}{2}\xi_0^2,\sqrt{2}\xi\right)},\;\frac{|\partial \epsilon_6/\partial \xi|}{\sqrt{2} N\left(\frac{1}{2}\xi_0^2,\sqrt{2}\xi\right)}\nb\\
&&\;\;\;\leq \frac{\kappa E\left(\frac{1}{2}\xi_0^2,\sqrt{2}\xi\right)}{\lambda } \Bigg\{\exp{\Big(\lambda \mathscr{V}_{0,\xi}(I)\Big)}-1\Bigg\}, ~~~~
\eqn
where $\lambda\equiv \kappa_0$,  and $I(\xi)$ is the corresponding error control function, given by
\bqn
\lb{FunctionI}
I(\xi)=\int^\xi |v|^{-1} \psi(v) dv.
\eqn
It is easy to get the error bounds for $\xi<0$ by replacing $\xi$ by $-\xi$ in the above expressions.

\subsection{Convergency of Error Control Functions $H(y)$ and $I(y)$}

In the above, we have introduced two error control functions $H(y)$ and $I(y)$,  defined by Eqs.(\ref{FunctionH}) and 
(\ref{FunctionI}),  in addition to $F(y)$. In this subsection, let us show that they are all convergent for the choice of
$g(y)$ and $q(y)$ given in Eq.(\ref{function}).
 
 To such goals, we first note that
\bqn
H(\xi)=F(y) \pm \frac{5}{16}\int \frac{d\xi}{ |\xi|^{5/2}},
\eqn
here  $\xi$ is defined by Eq.(\ref{xiy0}). Since $\xi\rightarrow +\infty$ when $y\rightarrow 0^+$, it follows that $H(\xi)$ and $F(y)$ either 
converge together or diverge together. With the choice of $g(y)$ and $q(y)$ in Eq.(\ref{function}) that satisfy the conditions (\ref{c1}) and (\ref{c2}), 
one immediately concludes that $H(\xi)$ is convergent as $y\rightarrow 0^+$. This is exactly one of the reasons we choose $g(y)$ and $q(y)$ to be 
given by  Eq.(\ref{function}).  

Similarly, as $y\rightarrow +\infty$, one can show that the error control function $I(\xi)$ defined in Eq.(\ref{FunctionI}) has the same convergency as
$F(y)$ for  the choice of $g(y)$ and $q(y)$ of Eq.(\ref{function}),  which satisfies the conditions (\ref{c3}).

\subsection{Matching the Individual Solutions}

So far, we have obtained the analytical approximate solutions near poles $y=0^{+},\; +\infty$, given by Eqs.(\ref{LG1}) and (\ref{LG2}), respectively, and in the neighborhoods of the turning 
points $y_{i}$, given, respectively, by Eqs.(\ref{airy solution1}) and (\ref{solutionW}). Now, we need to determine all the integration constants from the initial conditions by matching them 
on their boundaries (or  common regions) with the requirements that the mode function $\mu_{k}(\eta)$ and its first derivative $d\mu_{k}(\eta)/d\eta$ be continuous across each of their boundaries
(or in their common regions). 

We first note that   the approximate solutions (\ref{airy solution1}) around the turning point $y_0$ reduce to the LG solutions $\mu_k^{+}(y)$, as $y \rightarrow0^+$, while the approximate solutions (\ref{solutionW}) around the turning points $y_1$ and $y_2$ reduce to the LG solution $\mu_k^{-}(y)$, as   $y \rightarrow +\infty$.   

Assume that the universe was initially at the adiabatic vacuum \cite{Brandenberger2013CQG},
\bqn\lb{ini}
\lim_{y\rightarrow+ \infty}\mu_k(y)&=& \frac{1}{\sqrt{2\omega}}e^{-i\int \omega d\eta}\nb\\
&\simeq& \sqrt{\frac{k}{2}} \frac{1}{(-g)^{1/4}} \exp{\left(-i\int_{y_i}^y \sqrt{-g} dy\right)},\nb\\
\eqn
while the second condition shall be the Wronskian condition
\bqn
\mu_k(y) \mu_k^*(y)'-\mu_k^*(y) \mu_k(y)'=i.
\eqn
Applying the above conditions to the solution $\mu_k^{-}(y)$ given by Eq.(\ref{LG2}), we find that
\bqn\lb{cd-}
c_-=0,\;\;\;\;d_-=\sqrt{\frac{1}{2k}} e^{i\hat{\phi}},
\eqn
where $\hat{\phi}$ is an irrelevant phase factor. Without loss of the generality, we can always set it to zero. Similar considerations will be applied, when  other 
integration constants are concerned. 

On the other hand, to consider the matching between the solution $\mu_k^{-}(y)$ and the one given by Eq. (\ref{solutionW}), 
we note that for a positive and large $\xi$ the parabolic cylinder functions take the asymptotical forms \cite{Gil2004},
\bqn\lb{Wasymp}
W\left(\frac{1}{2}\xi_0^2,\sqrt{2}\xi\right) &\simeq& \left(\frac{2 j^2(\xi_0)}{\xi^2-\xi_0^2}\right)^{1/4} \cos{\mathfrak{D}},\nb\\
W\left(\frac{1}{2}\xi_0^2,-\sqrt{2}\xi\right) &\simeq& \left(\frac{2 j^{-2}(\xi_0)}{\xi^2-\xi_0^2}\right)^{1/4} \sin{\mathfrak{D}},
\eqn
where
\bqn
j(\xi_0)\equiv \sqrt{1+e^{\pi \xi_0^2}}-e^{\pi \xi_0^2/2},
\eqn
and
\bqn
\mathfrak{D}&\equiv& \frac{1}{2} \xi \sqrt{\xi^2-\xi_0^2}-\frac{1}{2}\xi_0^2 \ln{\left(\frac{\xi+\sqrt{\xi^2-\xi_0^2}}{|\xi_0|}\right)}\nb\\
&&+\frac{\pi}{4}+\phi\left(\frac{1}{2}\xi_0^2\right),
\eqn
with
\bqn\lb{phix}
\phi(x) &\equiv& \frac{x}{2}-\frac{x}{4} \ln{x^2}+\frac{1}{2} \text{ph}\Gamma\left(\frac{1}{2}+ix\right),
\eqn
where the phase  $\text{ph}\Gamma\left(\frac{1}{2}+i x\right)$ is zero when $x=0$, and determined by continuity, otherwise. 
Inserting the above into Eq.(\ref{solutionW}), and then comparing the resulting solution with $\mu_k^{-}(y)$, we find that the continuities of 
the mode function and its first derivatives with respect to $\eta$ yield,
\bqn\lb{aba}
\alpha_1&=& 2^{-3/4} k^{-1/2} j^{-1/2}(\xi_0),\nb\\
\beta_1&=& -i 2^{-3/4} k^{-1/2} j^{1/2}(\xi_0).
\eqn
 
To determine the coefficients $\alpha_0$ and $\beta_0$, let us consider the matching of the soolutions (\ref{solutionW}) and (\ref{airy solution1})
 in the region $y\in (y_0,y_1)$. In this region, $|y_0-y_1|$ is  large, as mentioned above, and $\xi$ is very negative. Thus, using the asymptotical 
 form (\ref{Wasymp}) of $W\left(\frac{1}{2}\xi_0^2,\sqrt{2}\xi\right)$, and the asymptotic form of the Airy functions,
\bqn\lb{negative large}
\text{Ai}(-x)&=&\frac{1}{\pi^{1/2} x^{1/4}}\cos{\left(\frac{2}{3}x^{3/2}-\frac{\pi}{4}\right)},\nb\\
\text{Bi}(-x)&=&-\frac{1}{\pi^{1/2} x^{1/4}}\sin{\left(\frac{2}{3}x^{3/2}-\frac{\pi}{4}\right)},
\eqn
 for $x \gg 1$, we find that the coefficients $\alpha_0$ and $\beta_0$ are given by,
\bqn\lb{coevv}
\alpha_0&=& \sqrt{\frac{\pi }{2k}}\; \left[j^{-1}(\xi_0) \sin{\mathfrak{B}} - i j(\xi_0) \cos{\mathfrak{B}}\right],\nb\\
\beta_0&=& \sqrt{\frac{\pi }{2k}}\; \left[j^{-1}(\xi_0) \cos{\mathfrak{B}} + i j(\xi_0) \sin{\mathfrak{B}}\right],
\eqn 
where 
\bq
\lb{functionb}
\mathfrak{B} \equiv \int_{y_0}^{y_1} \sqrt{-g}dy+\phi(\xi_0^2/2).
\eq

Finally, we consider the matching between $\mu^{+}_{k}(y)$ given by Eq.(\ref{LG1}) and the one given by Eq.(\ref{airy solution1}) in the region $y \in (0, y_0)$. 
It can be shown that the continuities of the mode function and its first derivatives with respect to $\eta$ yield
\bqn
\lb{cdp}
d_{+} &=& \frac{\alpha_0}{2 \sqrt{\pi}} \exp{\left(-\int_{0^+}^{y_0} \sqrt{g}dy\right)},\nb\\
c_{+} &=& \frac{\beta_0}{\sqrt{\pi}} \exp{\left(\int_{0^+}^{y_0} \sqrt{g}dy\right)}.
\eqn
The above completed the matching among these individual solutions given in each of the regions specified above. 
From such a process, it can be seen that all the integration constants are
uniquely determined form the initial conditions.  

Once we have determined these integration constants,  
let us now turn to consider some representative cases. In particular, in the case with three different single turning points, the numerical
(exact) and our  analytical approximate solutions are plotted in Fig. \ref{fig1}(a). The cases with two and one turning point(s) are plotted, respectively, in  Figs. \ref{fig1}(b) and \ref{fig1}(c).
From these figures, one can see clearly how well  the exact solutions are approximated by  our analytical ones.  
Many other cases have been also considered, and found that in all those cases the exact solutions are extremely well approximated 
by the analytical ones, thanks to the understanding of the error bounds, and the proper choice of the Liouville transformations.

\begin{figure}[t]
\centering
	{\includegraphics[width=75mm]{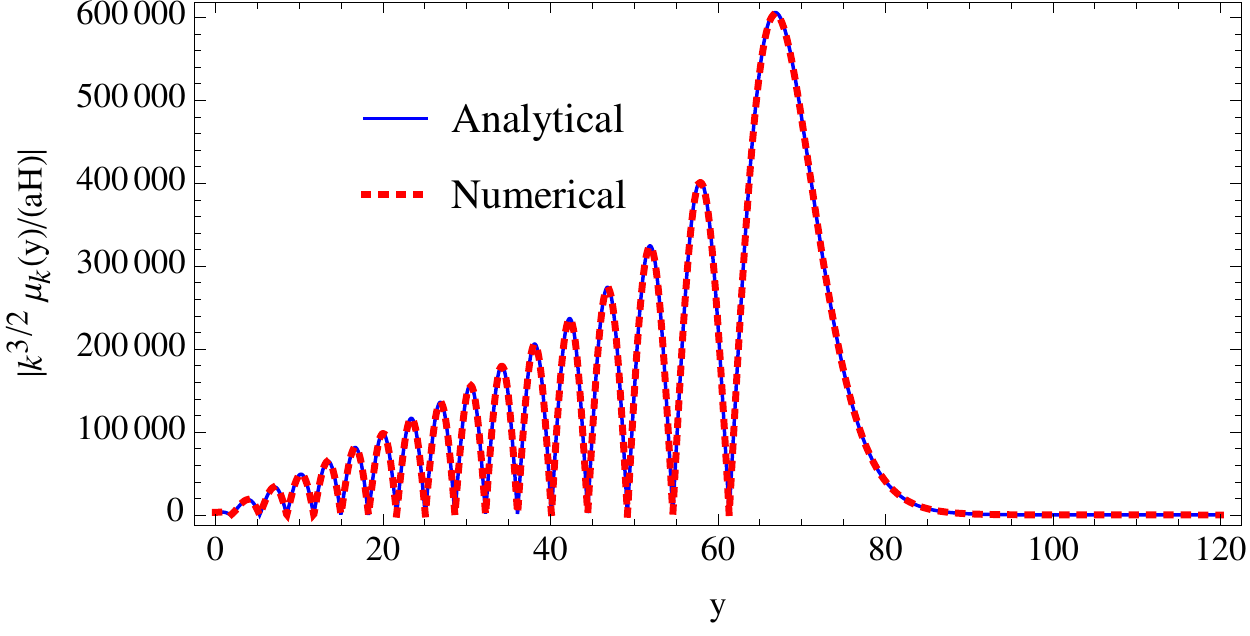}}
	{\includegraphics[width=75mm]{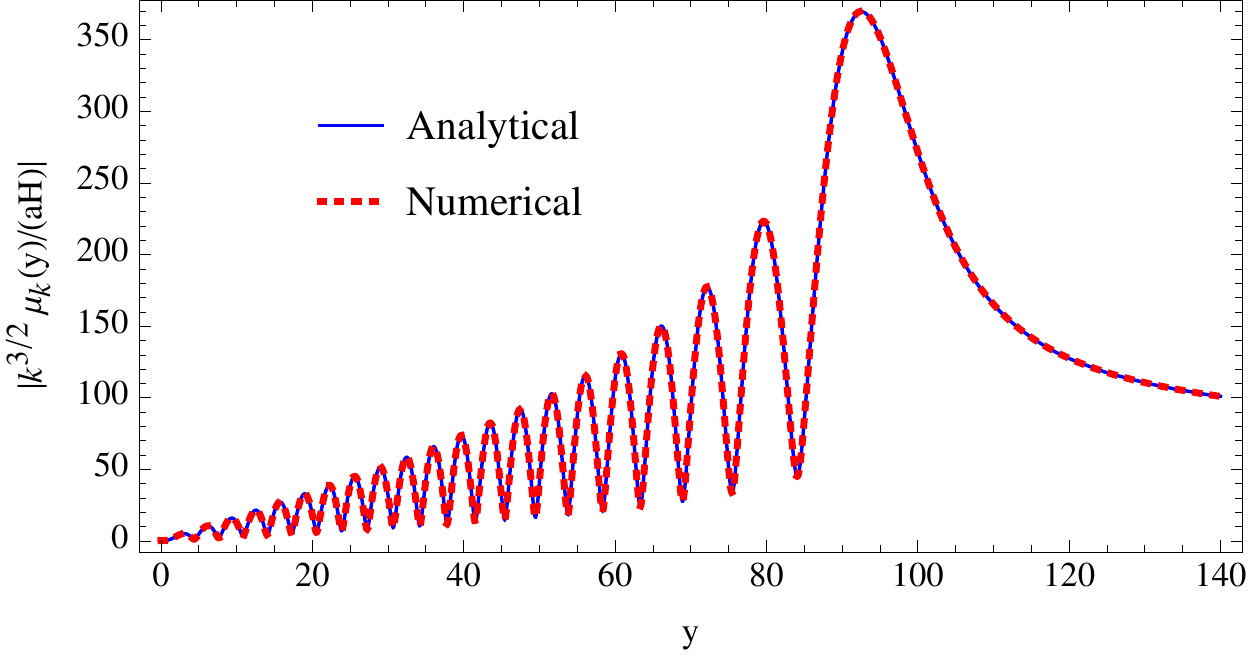}}
	{\includegraphics[width=75mm]{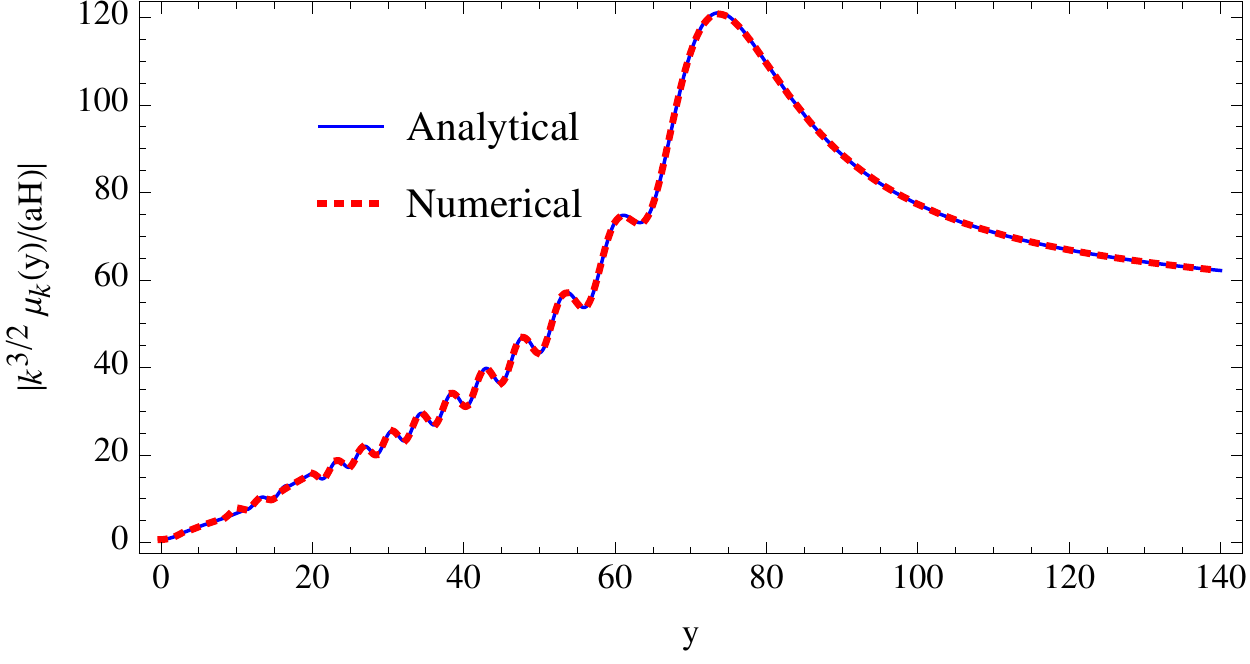}}
\caption{The numerical (exact)  (red dotted curves) and analytical (blue solid curves) solutions: 
(a) Top panel: Three  turning points  with  $b_1=3$, $b_2=2$. 
(b) Midlle panel: Two  turning points  with $b_1=2$, $b_2=1.00023$. 
(c) Low panel: One turning point  with $b_1=3.5$, $b_2=3.2$.
In all three cases, we have set $\nu=3/2$ and $\epsilon_*=0.01$.}
\lb{fig1}
\end{figure}

\section{Solutions for large $|\xi_0^2|$}
\renewcommand{\theequation}{3.\arabic{equation}} \setcounter{equation}{0}

In the last section,  the approximate analytical solutions around the turning point $y_0$ is expressed in terms of the Airy functions, 
while around the ones $y_1$ and $y_2$ they are  expressed in terms of the parabolic cylinder functions, which are uniform with 
respect to both real and purely imaginary values of $\xi_0$.  This difference comes from the fact that   Condition (b) holds at $y_0$,
while  in general only Condition (c) holds at $y_1$ and $y_2$.  
In this section, we consider the case where  $|\xi_0|$ is very large, so Condition 
(b) also holds  at both $y_1$ and $y_2$.  In the latter case,  we shall show that the approximate solutions around the turning points
$y_1$ and $y_2$ are also expressible in terms of the Airy functions.    

\subsection{Asymptotic Solutions for Real $y_{1,2}$} 

We first consider the case where  $y_1$ and $y_2$ are real. Similar to what we did around  the turning point $y_0$,  when Condition (b) is satisfied, 
the function $f^{(1)}(\xi)^2$ needs to have the same order of zeros as  $g(y)$ around the turning points $y_1$ and $y_2$. Thus, we can define 
a monotone increasing or decreasing function $\xi$ via the relations
\bqn
\lb{xidefine1}
f^{(1)}(\xi_i)^2=\pm\xi_i,
\eqn
where $\xi_i$ vanishes at the turning points $y_1$ and $y_2$, i.e., $\xi_1(y_1) = \xi_2(y_2)=0$. Note that, in order to distinguish them 
from the case considered in the last section, we replace the variable $\xi$ by $\xi_1$ ($\xi_2$) in the neighborhood of $y_1$ ($y_2$).   
Without loss of the generality, we can always choose $\xi_i$ to have the same sign as $g(y)$, and thus $\xi_1$ is a monotone increasing function around the
turning point $y_1$, while $\xi_2$ is a monotone decreasing function around the turning point $y_2$. Combining Eq.(\ref{Olver trans}) and Eq.(\ref{xidefine1}), 
we find that  
\bqn
\xi_1 &=& \begin{cases} \lb{xi11}
                   - \left(-\frac{3}{2} \int_{y_1}^{y} \sqrt{-g(y)} dy\right)^{\frac{2}{3}},&  y\in (y_0,y_1),\\ 
                   \left(\frac{3}{2} \int_{y_1}^{y} \sqrt{g(y)} dy\right)^{\frac{2}{3}},&  y\in (y_1,y_2) 
\end{cases} ~~~ \\
\xi_2 &=& \begin{cases} \lb{xi22}
                   - \left(\frac{3}{2} \int^y_{y_2} \sqrt{-g(y)} dy\right)^{\frac{2}{3}},&  y\in (y_2,\infty),\\ 
                   \left(-\frac{3}{2} \int^y_{y_2} \sqrt{g(y)} dy\right)^{\frac{2}{3}},&  y\in (y_1,y_2) .
\end{cases} 
\eqn
Then, Eq.(\ref{eomU}) reduces to  
\bqn
\frac{d^2U}{d\xi_i^2}=\big(\xi_i+\psi(\xi_i)\Big) U.
\eqn
Thus,  neglecting the $\psi(\xi_i)$ term as the first-order approximation, the above equation has the following general asymptotic solutions in terms of the Airy functions,
\bqn\lb{airy solution}
\mu^{(i)}_k(y)&\simeq& \hat{\alpha}_i  \left(\frac{\xi_i}{g(y)}\right)^{1/4}\text{Ai}(\xi_i) [1+\epsilon_3^{(i)}(\xi_i)]\nb\\
&&+\hat{\beta}_i \left(\frac{\xi_i}{g(y)}\right)^{1/4} \text{Bi}(\xi_i) [1+\epsilon_4^{(i)}(\xi_i)],
\eqn
here   
$\epsilon_1^{(i)}(\xi_i)$ denote the errors of the approximations, which are given  by similar expressions of Eqs.(\ref{error3}) and (\ref{error4}) for the turning point $y_0$,
so we shall not rewrite them here.  

Now let us turn to  determine  the integrations constants for the new solutions (\ref{airy solution}) near the turning points $y_1$ and $y_2$, which are valid only for 
$|\xi_0^2| \gg 1$, as we emphasized above.  First, the coefficients $c_-$ and $d_-$ are also given by  Eq.(\ref{cd-}), determined  from the initial conditions (\ref{ini}). 
To consider the matching between the solution $\mu_k^{-}(y)$ and $\mu_k^{(2)}(y)$, we note that $\xi_2$ is very large and negative. Thus, using the asymptotical form 
(\ref{negative large}), we find 
\bqn\lb{ab1}
\hat{\alpha}_2=\sqrt{\frac{\pi }{2k}},\;\;\
\hat{\beta_2}=i\sqrt{\frac{\pi }{2k}}.
\eqn
To determine the coefficients $\hat{\alpha}_1$ and $\hat{\beta}_1$, let us consider the matching of the solutions $\mu_k^{(2)}(y)$ and $\mu_k^{(1)}(y)$ in the region $y\in(y_1,y_2)$. 
Since  $|\xi_0| \gg 1$, we find that  both  $\xi_1$ and $\xi_2$ are very large in this region. Then, using the asymptotical form of the Airy functions
\bqn\lb{positive large}
\text{Ai}(x)&=&\frac{x^{-1/4}}{2\sqrt{\pi}}\text{ exp}\left(-\frac{2}{3}x^{3/2}\right).\nb\\
\text{Bi}(x)&=&\frac{x^{-1/4}}{\sqrt{\pi}}\text{ exp}\left(\frac{2}{3}x^{3/2}\right),
\eqn
we find that the coefficients $\hat{\alpha}_1$ and $\hat{\beta}_1$ are given by
\bqn\lb{ab2}
\hat{\alpha}_1&\simeq& 2 \beta_2 \exp{\left(\int_{y_{1}}^{y_{2}} \sqrt{g(y)}dy\right)} = i \sqrt{\frac{2 \pi}{ k}} e^{\mathfrak{A}},\nb\\
\hat{\beta}_1&\simeq& \frac{1}{2}\alpha_2 \exp{\left(-\int_{y_{1}}^{y_{2}} \sqrt{g(y)}dy\right)} = \sqrt{\frac{\pi }{8k}} e^{-\mathfrak{A}}, ~~~~
\eqn
where
\bqn
\mathfrak{A} \equiv \int_{y_1}^{y_2} \sqrt{g(y)}dy.
\eqn
In the region $y\in (y_0,y_1)$, we consider the matching of the solutions $\mu_k^{(1)}(y)$ and that  given by Eq.(\ref{airy solution}). In this region, $|y_0-y_1|$ is  large, as mentioned above, 
and  $\xi$ near  both of the turning points $y_0$ and $y_1$ is very negative. Thus,  using the asymptotical form (\ref{negative large}), we obtain
\bqn\lb{coe3}
\alpha_0 &=& \hat{\alpha}_1 \sin{\mathfrak{B}}+\hat{\beta}_1 \cos{\mathfrak{B}}\nb\\
&=&\sqrt{\frac{\pi }{2k}} \left(i 2 e^{\mathfrak{A}} \sin{\mathfrak{B}}+\frac{1}{2}e^{-\mathfrak{A}}\cos{\mathfrak{B}}\right),\nb\\
\beta_0 &=& \hat{\alpha}_1 \cos{\mathfrak{B}}-\hat{\beta}_1 \sin{\mathfrak{B}}\nb\\
&=&\sqrt{\frac{\pi }{2k}} \left(i 2 e^{\mathfrak{A}} \cos{\mathfrak{B}}-\frac{1}{2}e^{-\mathfrak{A}}\sin{\mathfrak{B}}\right),
\eqn
where for large $\xi_0$ we have $\phi(\xi_0^2/2)=0$, and $\mathfrak{B}$ is given by Eq.(\ref{functionb}).

\subsection{Asymptotic Solutions for Complex $y_{1,2}$} 

When $y_{1,2}$ are complex, there is only one single real turning point $y_0$. When  $|\xi_0|$ is large, one finds that Condition (a) and (b) specified in Section II.B
hold in the whole region $y\in (0^{+},+\infty)$. Thus, in this case the approximate solutions (\ref{airy solution1}) are valid in the region  $y\in (0^+,\infty)$. This  
reduces  exactly to the cases studied by Wang and  Yamamoto, Kobayashi, and Nakamura  in  \cite{HL}. Comparing the approximate solutions (\ref{airy solution1}) 
with the initial adiabatic vacuum 
condition (\ref{ini}), and considering  the asymptotic form (\ref{negative large}), we find that   the coefficients $\alpha_0$ and $\beta_0$ are given by
\bqn
\alpha_0=\sqrt{\frac{\pi}{2k}},\;\;\;\beta_0=i \sqrt{\frac{\pi}{2k}}.
\eqn

Finally, we note that the above results can be also obtained by directly expand the parabolic cylinder functions in  terms of the Airy functions when  $|\xi_0|$ is large.
For details, see Appendix D.

\section{Power Spectra and Indices of Scalar and Tensor Perturbations}
\renewcommand{\theequation}{4.\arabic{equation}} \setcounter{equation}{0}


In the above sections, we constructed the approximate analytical solutions of the linear perturbations (\ref{eom}) for scalar, vector or tensor fields in the framework  of slow-roll inflation.
In GR and  Ho\v{r}ava-Lifshitz gravity, vector perturbations decay rapidly with the expansion of the universe and have negligible effects on observations  \cite{InfGR,InfHL}. So, in this section 
we consider only the scalar and tensor perturbations, and pay particular attention on the modifications of them  due to the quantum effects. 

As mentioned previously, although the linear perturbations of all the types can be written in the form (\ref{eom}), the constants $b_i$ and the function $z$ are different for different types of
perturbations. from which it can be shown that  $z$  for the scalar and tensor perturbations is given, respectively,  by \cite{InfGR}, 
\bqn
\lb{zfunction}
z_s = \sqrt{2\epsilon}\; a, \;\;\;
z_T = a,
\eqn
where we use the subscripts $s$ and $T$ to denote the scalar and tensor perturbations, respectively. 
Recall that the slow-roll parameter $\epsilon$ is defined as $\epsilon = - \dot{H}/H^2$.  Then, the corresponding 
$\nu_{s}$ and $\nu_{T}$ are given by Eq.(\ref{ab}). To the first-order  
of the slow-roll parameters, we have \cite{uniformPRL}
\bq
\lb{zszt}
\nu_{s} \simeq  \frac{3}{2} + 2\epsilon + \delta_1,\;\;\;
\nu_{T} =  \frac{3}{2} + \epsilon,
\eq
with $\delta_1 \equiv \ddot{\phi}/H\dot{\phi}$, and $\phi$ denotes the inflaton. Similarly, we shall use $b^{s}_{i}$ and $b^{T}_{i}$ to denote, respectively,
the  scalar and tensor perturbations,  $\mu_{k}^{s}(y, b^{s}_i, \nu_{s})$ and $\mu_{k}^{T}(y, b^{T}_i, \nu_{T})$ the 
corresponding mode functions. With the above in mind, in the following we shall ignore the subscripts whenever it is
possible.  
 
 \subsection{Power Spectra}

To calculate the power spectra,    we only need to consider the limit $y\to 0^+$. Then, from  Eq.(\ref{positive large})
we can see that in this limit only  the growing  mode is relevant, so we have   
\bqn
\mu_k(y) \simeq \beta_0 \left(\frac{1}{\pi^2 g(y)}\right)^{1/4} \text{exp}\left(\int_{y}^{y_0} dy \sqrt{g(y)}\right),
\eqn
where $\beta_0$ is given by Eq. (\ref{coevv}).
Then,  the corresponding  power spectrum is given by,  
\bqn
\lb{power spectrum}
\Delta^2(k) &\equiv&\frac{k^3}{2\pi^2} \left|\frac{\mu_k(y)}{z}\right|^2\nb\\
&= & \left(\frac{k^2y}{4\pi^2 z^2 \nu}\right)  {\cal{A}}\exp{\left(2\int_{y}^{y_0} \sqrt{g(y)}dy\right)},
\eqn
where
\bqn\lb{Q}
{\cal{A}} &\equiv& \frac{2k |\beta_0|^2}{\pi}\nb\\
&=&1+2 e^{\pi \xi_0^2} + 2 e^{\pi\xi_0^2/2} \sqrt{1+e^{\pi\xi_0^2}}\; \cos{2\mathfrak{B}},
\eqn
with $\xi_0^2$ being  given by
\bqn
\lb{xi02}
\xi_0^2=\pm\left| \frac{2}{\pi} \int_{y_1}^{y_2} \sqrt{\pm g(y)} dy\right|,
\eqn
where ``$+$'' corresponds to the case where $y_1$ and $y_2$ are real, and  ``$-$" to the case where $y_1$ and $y_2$ are complex conjugate 
 $y_1=y_2^{*}$. 
The factor  $\mathfrak{B}$ is given by Eq.(\ref{functionb}). The integrals in $\xi_0^2$ and $\mathfrak{B}$ can be carried out explicitly. Let us
 first consider the integral (\ref{xi02}). For the case where the equation $g(y)=0$ has three real and different roots, we  find
 \bqn
 \lb{4a.1}
\xi_0^2&=&\frac{2\sqrt{b_2}\epsilon_*^2}{\pi} \int_{y_1}^{y_2} \sqrt{\frac{(y^2-y_0^2)(y^2-y_1^2)(y_2^2-y^2)}{y^2}}dy\nb\\
&=& - \frac{2\sqrt{b_2}\epsilon_*^2}{3 \pi \sqrt{y_2^2-y_0^2}} \nb\\
&& \times \Bigg\{(y_0^2-y_2^2)(y_0^2+y_1^2+y_2^2) E\left(\frac{y_2^2-y_1^2}{y_2^2-y_0^2}\right)\nb\\
&&+(-y_0^4+2 y_1^2 y_2^2+y_0^2y_2^2+y_0^2y_1^2) K\left(\frac{y_2^2-y_1^2}{y_2^2-y_0^2}\right)\nb\\
&&- 3 y_0^2y_1^2 \Pi\left(\frac{y_2^2-y_1^2}{y_2^2},\frac{y_2^2-y_1^2}{y_2^2-y_0^2}\right)\Bigg\},
\eqn
where  $E(x)$, $K(x)$, $\Pi(x,y)$ denote the Elliptic integrals \cite{AS72}, and are given explicitly in Appendix E. It is easy to show that the above expression is also valid for the case 
where $y_1$ and $y_2$ are complex conjugate, for which  $\xi_0^2$ is  negative. In addition, in the case where   the equation $g(y)=0$ has one single real root
$y_0$ and one double real root $y_1(=y_2)$, we have $\xi_0^2=0$.  On the other hand,  we also find
\begin{widetext}
\bqn
&&\mathfrak{B}=\int_{y_0}^{y_R} \sqrt{-g(y)} dy+\phi\left(\frac{1}{2} \xi_0^2\right)\nb\\
&&\;\;\;\;\;=\frac{\sqrt{b_2}\epsilon_*^2}{3} \sqrt{(y_R^2-y_0^2)(y_R^2-y_1^2)(y_R^2-y_2^2)}\Bigg\{1+\sqrt{\frac{y_2^2-y_R^2}{y_2^2-y_0^2}} \frac{1}{(y_0^2-y_1^2)(y_2^2-y_R^2) } \sqrt{\frac{(y_1^2-y_0^2)^2}{(y_1^2-y_R^2)(y_R^2-y_0^2)}}\nb\\
&&\;\;\;\;\;\;\;\;\;\;\;\;\;\;\;\;\;\;\times \Bigg[(y_0^2-y_2^2) (y_0^2+y_1^2+y_2^2) E\left(\arcsin\left[{\sqrt{\frac{y_R^2-y_0^2}{y_1^2-y_0^2}}}\right],\frac{y_0^2-y_1^2}{y_0^2-y_2^2}\right)\nb\\
&&\;\;\;\;\;\;\;\;\;\;\;\;\;\;\;\;\;\;\;\;\;\;\;\;+\left(y_2^4-2 y_0^2 y_1^2 -y_0^2 y_2^2-y_1^2 y_2^2 \right) F\left(\arcsin\left[{\sqrt{\frac{y_R^2-y_0^2}{y_1^2-y_0^2}}}\right],\frac{y_0^2-y_1^2}{y_0^2-y_2^2}\right)\nb\\
&&\;\;\;\;\;\;\;\;\;\;\;\;\;\;\;\;\;\;\;\;\;\;\;\;+3 y_1^2 y_2^2 \Pi \left(1-\frac{y_1^2}{y_0^2}, \arcsin\left[{\sqrt{\frac{y_R^2-y_0^2}{y_1^2-y_0^2}}}\right],\frac{y_0^2-y_1^2}{y_0^2-y_2^2}\right)\Bigg]\Bigg\}+\phi\left(\frac{1}{2} \xi_0^2\right),
\eqn
\end{widetext}
where $y_R=\text{Re}(y_1)$ when $y_1$ is complex, and $y_R=y_1$ when $y_1$ is real. The functions $E(x,y)$, $F(x,y)$, and $\Pi(x,y)$ denote the other types of the Elliptic integrals \cite{AS72}, and 
are given explicitly in Appendix E. The phase $\phi\left(\frac{1}{2}\xi_0^2\right)$ is given by Eq. (\ref{phix}). When $y_1$ and $y_2$ are real, the above expression can be simplified into the form
\bqn
\mathfrak{B}&=&\frac{\sqrt{b_2} \epsilon_*^2}{3 \sqrt{y_2^2-y_0^2}} \Bigg\{(y_2^2-y_0^2) (y_0^2+y_1^2+y_2^2) E\left(\frac{y_0^2-y_1^2}{y_0^2-y_2^2}\right)\nb\\
&&\;\;\;\;+\left(-y_2^4+2 y_0^2 y_1^2 +y_0^2 y_2^2+y_1^2 y_2^2 \right) K\left(\frac{y_0^2-y_1^2}{y_0^2-y_2^2}\right)\nb\\
&&\;\;\;-3 y_1^2 y_2^2 \Pi \left(1-\frac{y_1^2}{y_0^2},\frac{y_0^2-y_1^2}{y_0^2-y_2^2}\right)\Bigg\}+\phi\left(\frac{1}{2}\xi_0^2\right).\nb\\
\eqn
In addition, if  we have   $y_1=y_2$, then we find 
\bqn
\mathfrak{B}&=&\sqrt{b_2} \epsilon_*^2 \Bigg\{\frac{\sqrt{y_1^2-y_0^2}}{3} (y_0^2+2 y_1^2) -\frac{\pi}{2} y_0 y_1^2 \nb\\
&&+\arctan{\left(\frac{y_0}{\sqrt{y_1^2-y_0^2}}\right)}\Bigg\},\; (y_1 = y_2). ~~~~~~
\eqn

To process further, let us first write $g(y)$ in the form, 
\bqn
g(y)=\frac{y_0^2-y^2}{y^2} (h_0+h_1 y^2+h_2 y^4),
\eqn
where
\bqn
h_0&=& b_2 \epsilon_*^4 y_0^4-b_1 \epsilon_*^2 y_0^2+1, \nb\\ 
h_1&=& b_2 \epsilon_*^4 y_0^2-b_1 \epsilon_*^2,\;\;\;
h_2 = b_2 \epsilon_*^4,
\eqn
and  
\bqn
&&\sqrt{h_0+h_1y^2+h_2 y^4} = 1-\frac{b_1}{2}  (y^2+ y_0^2) \epsilon_*^2\nb\\
&& ~~~~ - \Big[\frac{b_1^2}{8}\left(y^2+y_0^2\right)^2 -\frac{b_2}{2} \left(y^4+y^2 y_0^2+y_0^4\right)\Big]\epsilon_*^4 \nb\\
&& ~~~~ +{\cal{O}}\left(\epsilon_*^6\right).
\eqn
Then, we obtain
\bqn
&&\int_y^{y_0} \sqrt{g(\hat{y})}d\hat{y} = -y_0 z_0 \ln{\frac{y}{y_0+\sqrt{y_0^2-y^2}}}\nb\\
&&\;\;\;\;+\frac{(y_0^2-y^2)^{3/2}}{15} (5z_1+3 z_2 y^2+2z_2 y_0^2)\nb\\
&&~~~~~  -z_0 \sqrt{y_0^2-y^2} +{\cal{O}}\left(\epsilon_*^6\right),
\eqn
where
\bqn
z_0&=&1-\frac{b_1}{2}y_0^2 \epsilon_*^2+\frac{4 b_2-b_1^2}{8}y_0^4 \epsilon_*^4,\nb\\
z_1&=& -\frac{b_1}{2}\epsilon_*^2+\frac{4b_2-2b_1}{8}y_0^2 \epsilon_*^4,\nb\\
z_2&=& \frac{4 b_2-b_1^2}{8} \epsilon_*^4.
\eqn

On the other hand, from Eq.(\ref{roots}) we find
 \bqn
 y_0\simeq \nu+\delta_1 \epsilon_*^2+\delta_2 \epsilon_*^4+{\cal{O}}(\epsilon_*^6),
 \eqn 
 where
 \bqn
 \delta_1=\frac{b_1 \nu^3}{2},\;\;\;
\delta_2= \frac{1}{8} (7 b_1^2-4 b_2) \nu^5. 
 \eqn
 Inserting the above expressions into Eq.(\ref{power spectrum}), we obtain
 \bqn
 \lb{spectra}
 \Delta^2(k) &=& \Delta_{0}^{2}(k)  {\cal{Q}}, 
 \eqn
 where
 \bqn\lb{pow}
\Delta^{2}_0&\equiv &  \left(\frac{k^2}{4\pi^2 z^2 \nu} \right) y^{1-2\nu} \left(\frac{2 \nu}{e}\right)^{2\nu},
\eqn
represents the power spectra of GR (for $b_1=0=b_2$),   obtained in \cite{uniformPRL}.  
The function ${\cal{Q}}$ is defined as
\bqn
  \lb{Q1}
 {\cal{Q}}&\equiv& {\cal{A}}\; \Bigg\{1+\frac{5}{6}b_1 \nu^3 \epsilon_*^2+\frac{\nu^5}{360} \Big(-30 b_1+444b_1^2\nb\\
 &&\;-275b_2+125 b_1^2\nu\Big)\epsilon_*^4+{\cal{O}}\left(\epsilon_*^6\right)\Bigg\}.
 \eqn
 
\subsection{The Spectral Indices}

The corresponding scale and tensor spectral indices can be calculated directly from the power spectra (\ref{power spectrum}), which are given, respectively, by  
\bqn
n_s &\equiv& 1+ \frac{d\ln{\Delta^2_{s}}}{d\ln{k}}\nb\\
&=& 4 -2  \int_y^{y_0} \frac{1-2 b_1^s \epsilon_*^2 \hat{y}^2+3 b_2^s \epsilon_*^4 \hat{y}^4}{\sqrt{g_s(\hat{y})}}d\hat{y}\nb\\
&=& \left. 4-2 \sqrt{\nu^2_s-y^2+b_1^s \epsilon_*^2 y^4 -b_2^s \epsilon_*^4 y^6}\right|_{y = 0}\nb\\
&=& 4-2 \nu_s\nb\\
&=& n_s^{\text{GR}},\\
n_{T}  &\equiv& \frac{d\ln{\Delta_T^2}}{d\ln{k}}\nb\\
&=& 3 -2  \int_y^{y_0} \frac{1-2 b_1^T \epsilon_*^2 \hat{y}^2+3 b_2^T \epsilon_*^4 \hat{y}^4}{\sqrt{g_{T}(\hat{y})}}d\hat{y},\nb\\
&=& \left. 3-2 \sqrt{\nu_T^2-y^2+b^T_1 \epsilon_*^2 y^4 -b^T_2 \epsilon_*^4 y^6}\right|_{y = 0}\nb\\
&=& 3-2 \nu_T\nb\\
&=& n_T^{\text{GR}},
\eqn
with $g_{I}({y}) \equiv g(y, b^{I}_{i}, \nu_{I})\; (I = s, T)$. That is, {\em to the first-order approximations of the slow-roll parameters,  
the power spectrum indices of scalar and tensor perturbations are the same as those given in GR}.
This is an important conclusion and  different from the results obtained initially in the first reference of \cite{Martin2001}, 
but similar to the ones obtained later in \cite{ACD}, in which only the
term $b_1$ was considered, and then the corresponding mode function can be obtained analytically.

It should be noted that in the above expressions we have assumed that the quantities $\nu$, $b_1$ and $b_2$ are all constants. 
This assumption is correct if one only considers the first-order slow-roll approximations. In this case, the quantity ${\cal{A}}$ does not 
depend on $k$, and, as a result,  the spectral indices $n_{s,T}$ do not receive any corrections from ${\cal{A}}$.  
To consider the quantum effects of the high-order derivative terms on $n_{s,T}$, we need to 
go beyond the first-order approximations or consider other backgrounds of the universe. However, 
the factor ${\cal{A}}$ does affect the amplitudes of the power spectra $\Delta^{2}_{s, T}(k)$, as one can see from Eqs.(\ref{spectra}) and (\ref{Q1}). 

 \begin{widetext}
 
\begin{figure}[t]
\centering
	{\includegraphics[width=75mm]{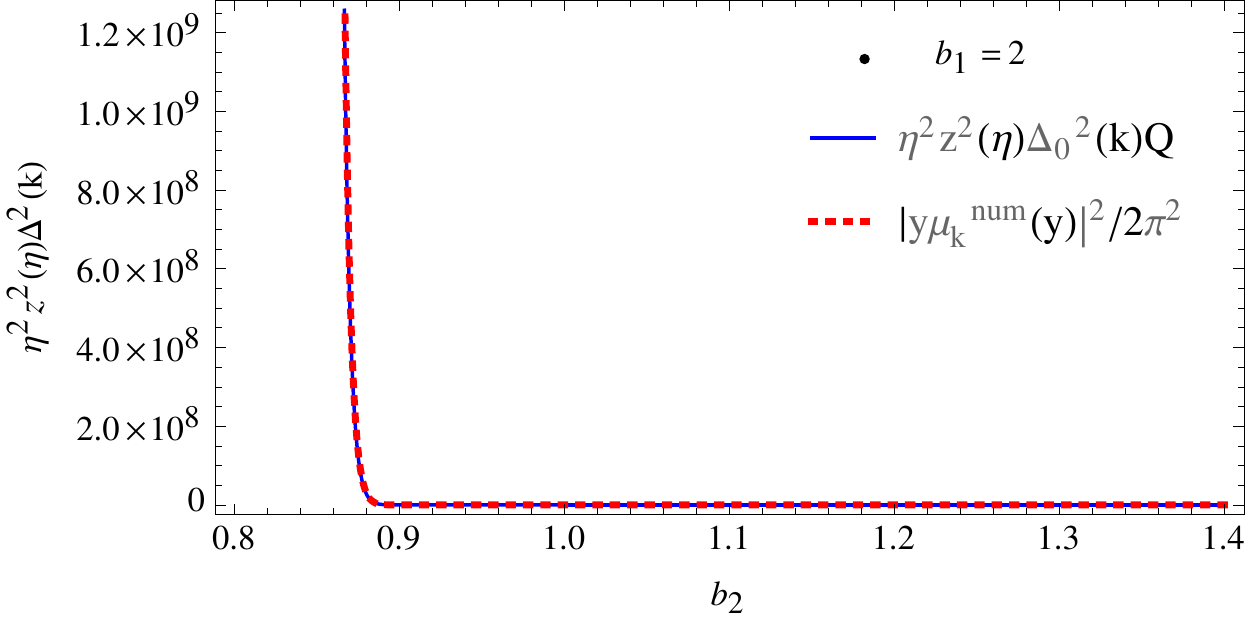}}
        	{\includegraphics[width=75mm]{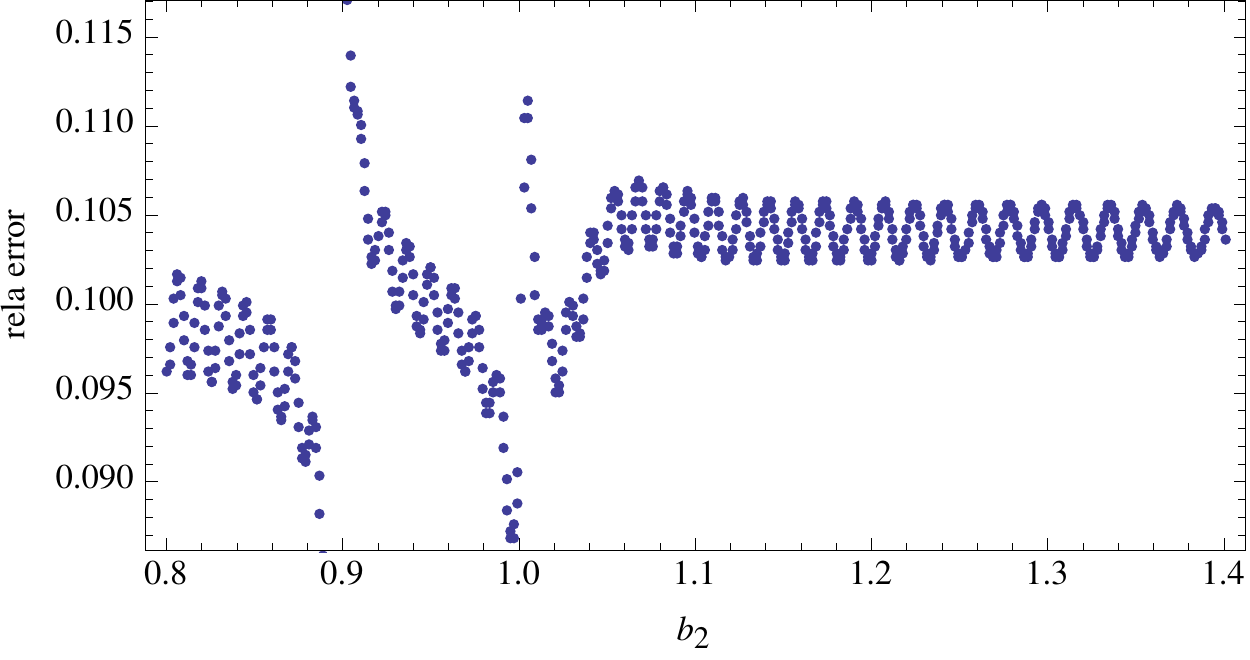}}
        	{\includegraphics[width=75mm]{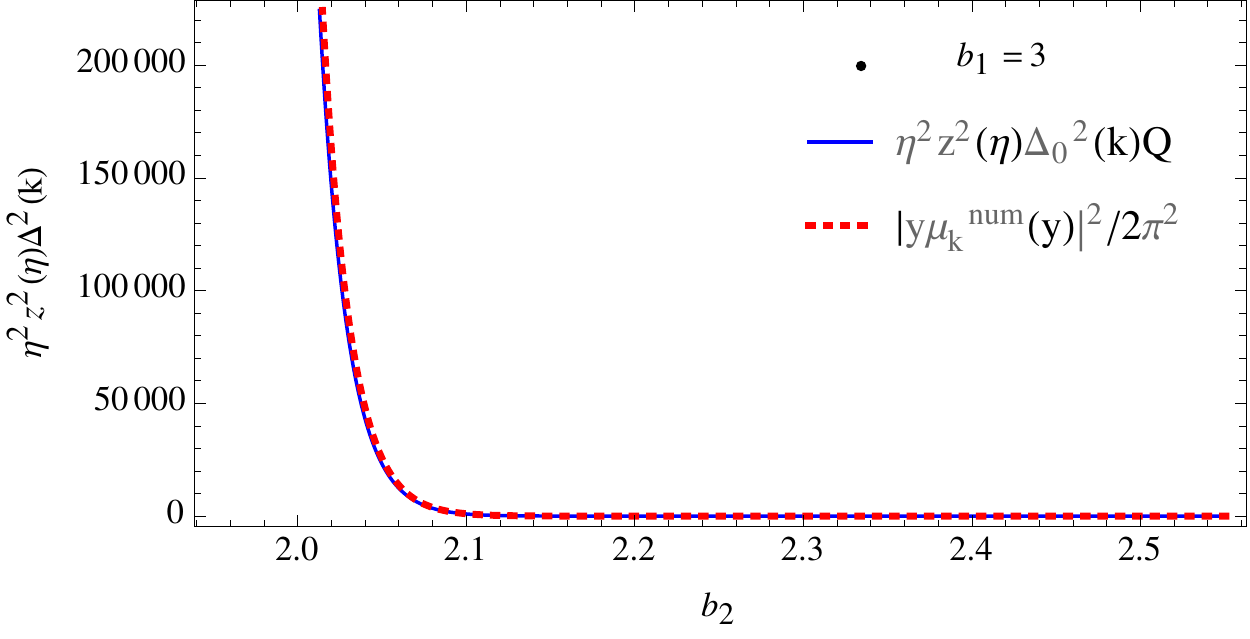}}
        	{\includegraphics[width=75mm]{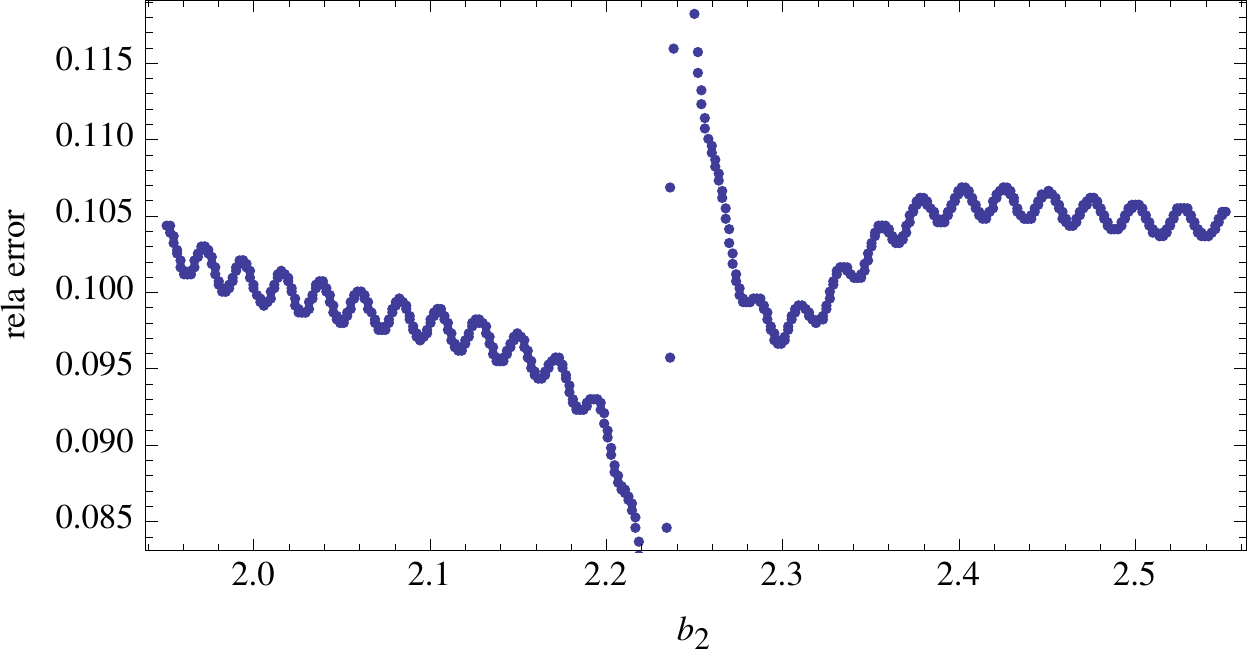}}
	 {\includegraphics[width=75mm]{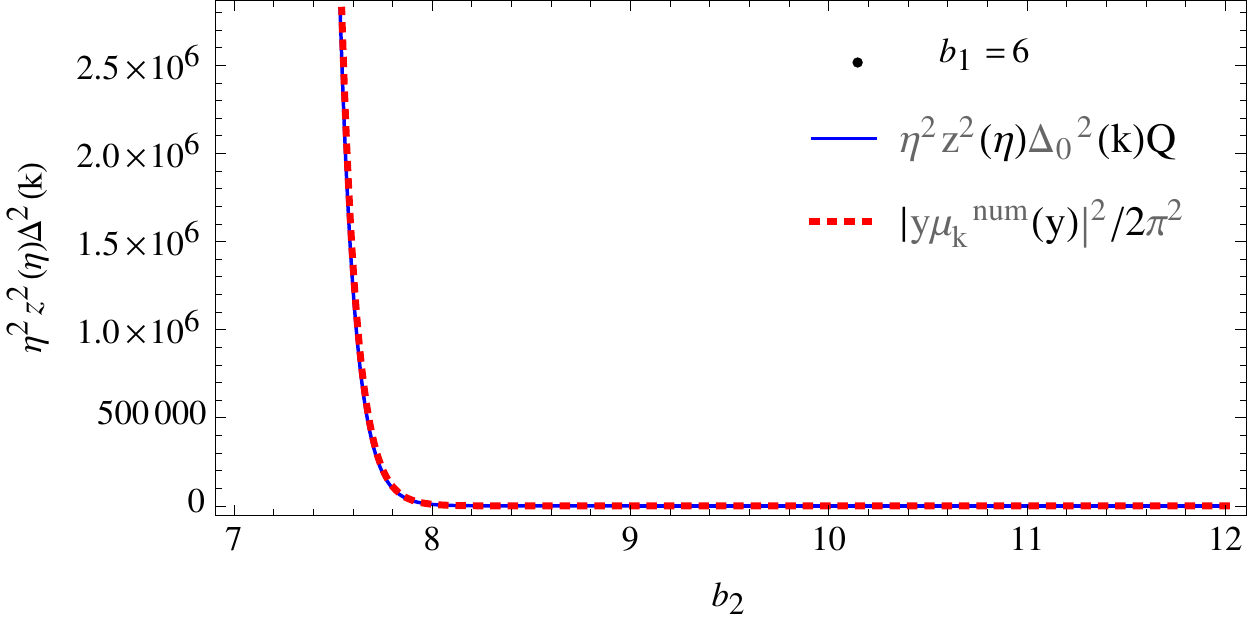}}
        	{\includegraphics[width=75mm]{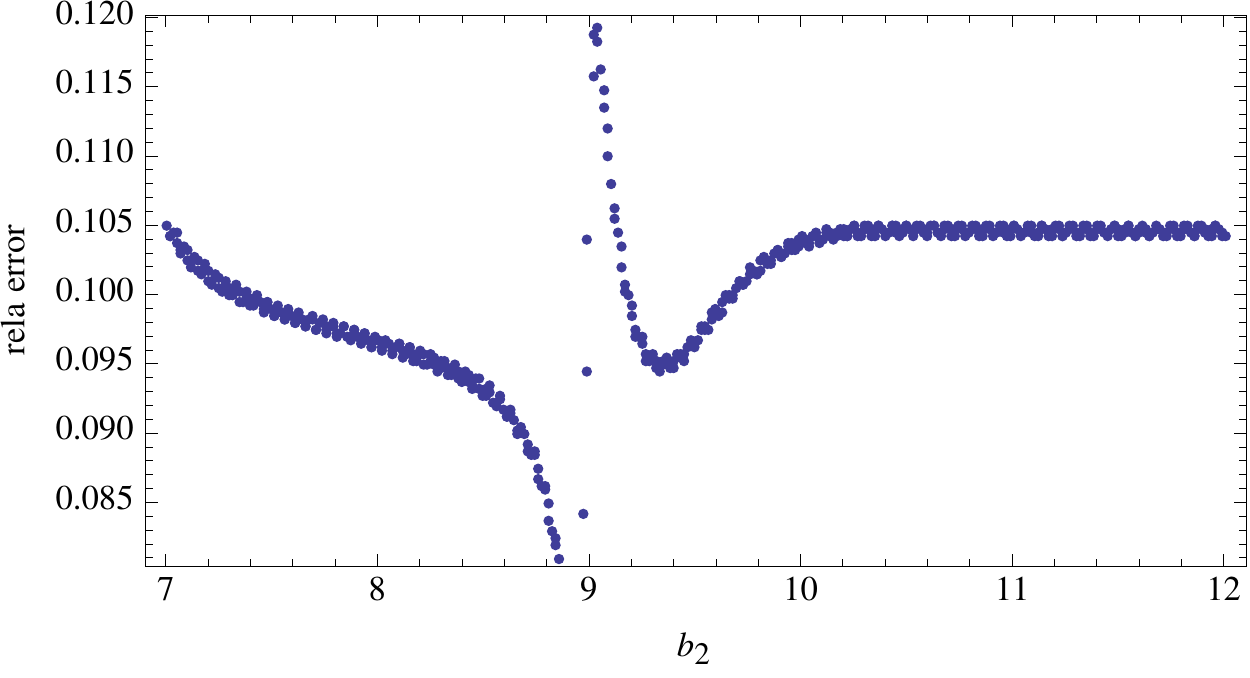}}
\caption{The numerical (exact)  (red dotted curves) solution of $\eta^2z^2 \Delta_{\text{num}}^2(k)$, the  
analytical (blue solid curves) one $\eta^2z^2 \Delta_0^2(k){\cal{Q}}$, and their relative errors defined by Eq.(\ref{relativeerror}) vs the free parameter $b_2$ for
 $\nu=3/2$, $\epsilon_*=0.01$. Top panel: $b_1=2$. Middle Panel: $b_1=3$. Low Panel: $b_1=6$.}
\lb{fig3}
\end{figure}

\begin{figure}[t]
\centering
	{\includegraphics[width=75mm]{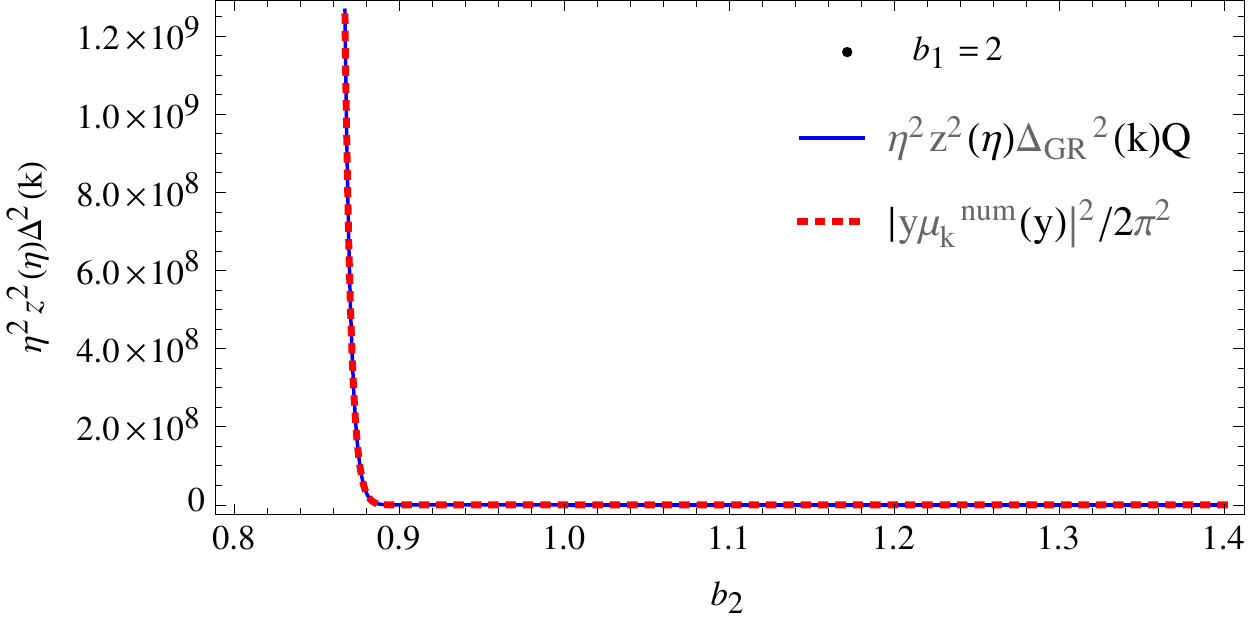}}
        	{\includegraphics[width=75mm]{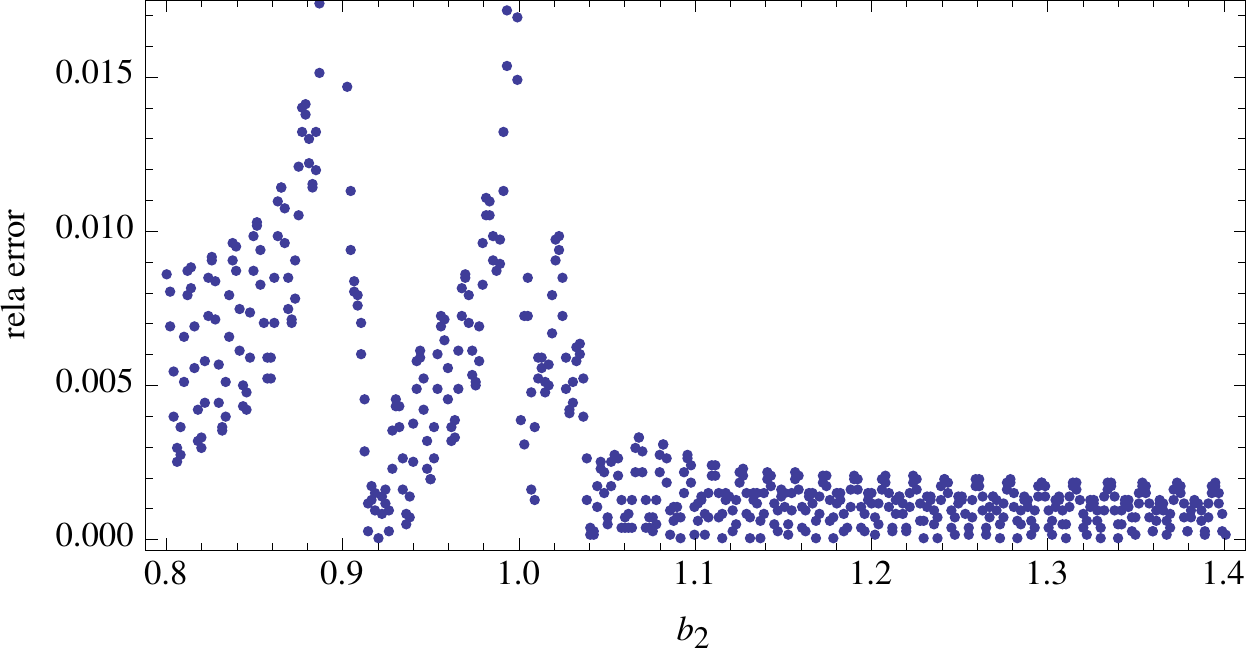}}
        	{\includegraphics[width=75mm]{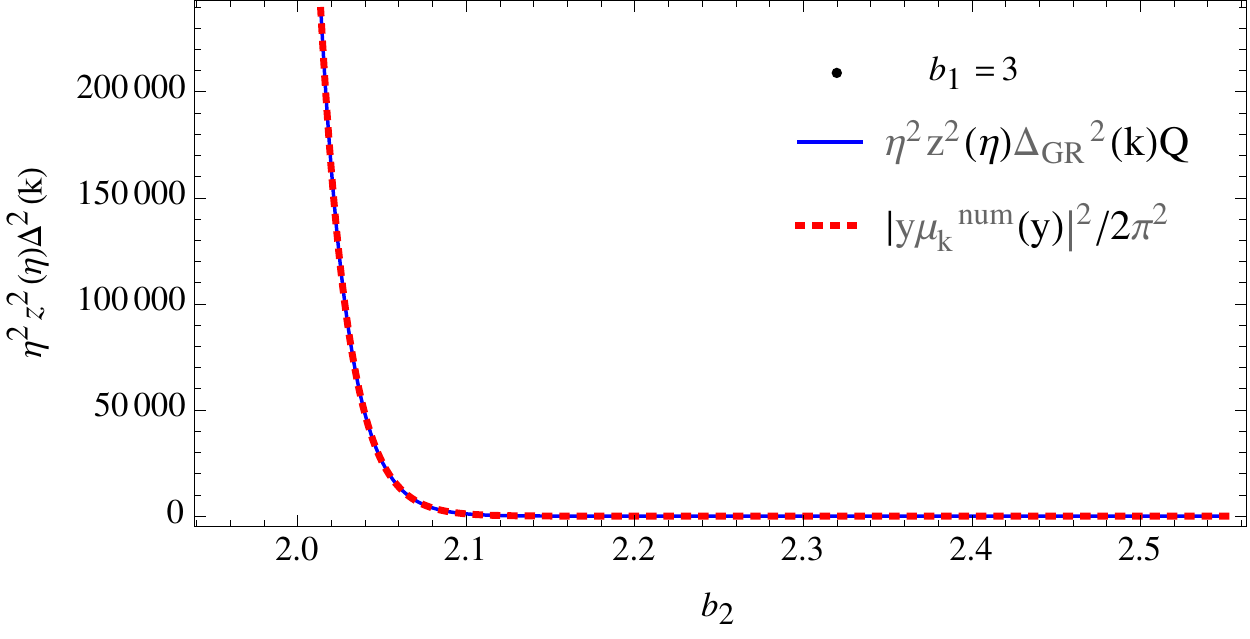}}
        	{\includegraphics[width=75mm]{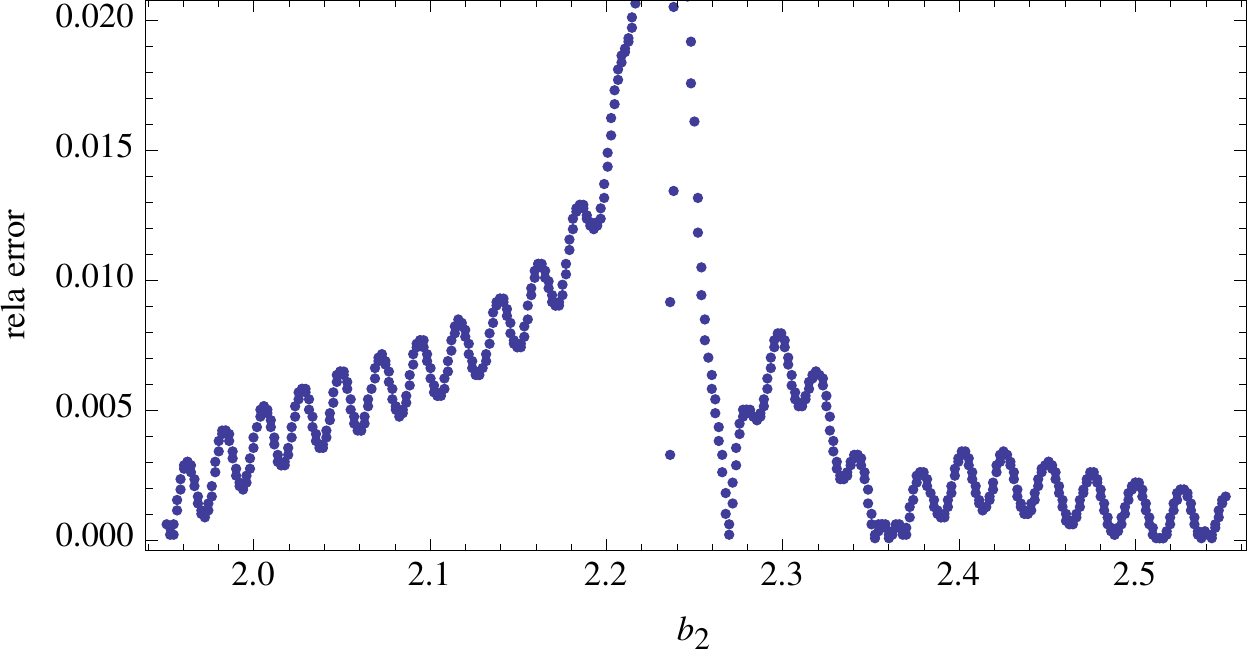}}
	 {\includegraphics[width=75mm]{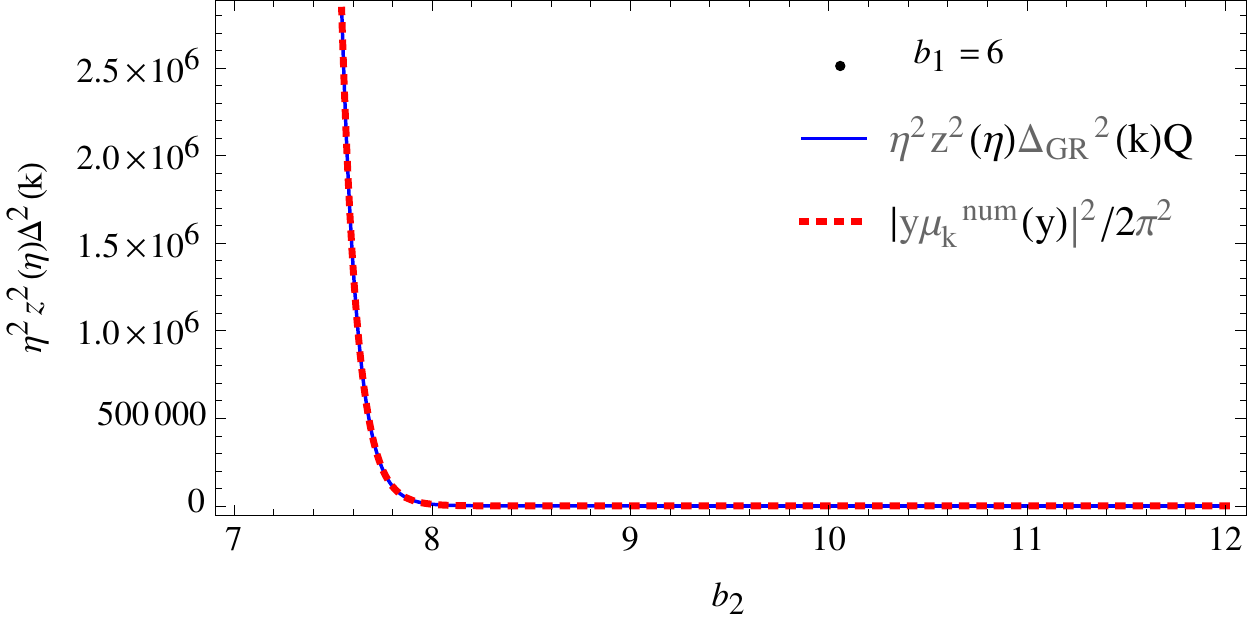}}
        	{\includegraphics[width=75mm]{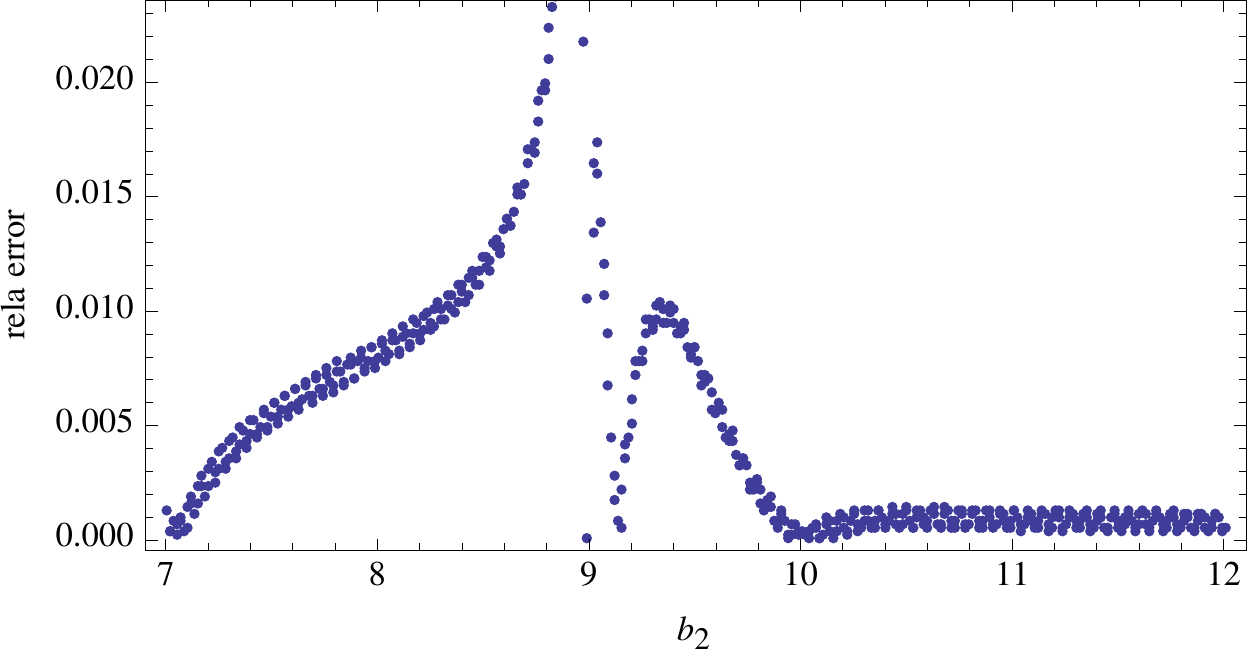}}
\caption{The numerical (exact)  (red dotted curves) solution of $\eta^2z^2 \Delta_{\text{num}}^2(k)$,  the  
analytical (blue solid curves) one $\eta^2z^2 \Delta_{\text{GR}}^2(k){\cal{Q}}$, and their relative errors defined by Eq.(\ref{relativeerror}) vs the free parameter $b_2$ for
 $\nu=3/2$, $\epsilon_*=0.01$. Top panel: $b_1=2$. Middle Panel: $b_1=3$. Low Panel: $b_1=6$.}
\lb{fig4}
\end{figure}
\end{widetext}

\subsection{Error Analysis}

In the last  subsections, we have obtained the explicit expressions for the power spectra and their spectral indices. In order to show the corresponding errors, 
in this subsection, we compare them with the numerical (exact) results. The numerical solutions $\mu^{\text{num}}_k(y)$ of Eq. (\ref{eom58}) with the initial conditions
 (\ref{ini}) can be easily obtained by the numerical integrations. Then, we find that
 
\bqn
\lb{relativeerror}
&& \left|\frac{\Delta^2(k)-\Delta^2_{\text{num}}(k)}{\Delta^2_{\text{num}}}\right|\nb\\
&& ~~~~~~~~~~ = \left|\frac{\eta^2z^2\Delta^2(k)-|y\mu_k^{\text{num}}(y)|/(2\pi^2)}{|y\mu_k^{\text{num}}(y)|/(2\pi^2)} \right|,~~~~
\eqn
where
\bqn
\Delta_{\text{num}}^2(k)\equiv \frac{k^3}{2\pi^2}\left|\frac{\mu_k^{\text{num}}(y)}{z(\eta)}\right|^2.
\eqn
In Fig.  \ref{fig3} we plot out the analytical and numerical values of $\eta^2z^2 \Delta^2(k)$ vs  $b_2$, and their relative errors, from which it can be seen  that for most of the cases, the relative errors are
about $10\%$, which is approximately in the same level as that obtained in GR by using the uniform approximation in the first-order approximations \cite{uniformPRL}. 

When $b_1=0=b_2$, the power spectra in Eq. (\ref{pow}) reduce to the usual results $\Delta_0^2(k)$, whose relative error is well understood by comparing it with the exact analytical result $\Delta_{\text{GR}}^2(k)$, i.e.,
\bqn
\Delta_{\text{GR}}^2(k)=\Delta_0^2(k) (1+\delta_e),
\eqn
where $\delta_e$ is in the level of $10\%$. Thus, the relative errors of the power spectra (\ref{pow}) can be divided into two parts, one  comes from $\Delta^2_0(k)$ and  the other comes from 
${\cal{Q}}$, 
\bqn
\Delta_{\text{exact}}^2(k)=\Delta_0^2(k) (1+\delta_e) {\cal{Q}} (1+\delta_{{\cal{Q}}}),
\eqn
where $\Delta_{\text{exact}}$ represents the exact power spectra. Since $\delta_e$ is well understood in GR,  
to improved the uniform approximations, one can  replace the $\Delta_0^2(k)$ in Eq. (\ref{pow}) by  $\Delta_{\text{GR}}^2(k)$, i.e., 
\bqn
\Delta_{\text{exact}}^2(k) &=&\Delta_{\text{GR}}^2(k) {\cal{Q}} (1+\delta_{{\cal{Q}}})\nb\\
&=&\Delta^2(k) (1+\delta_{{\cal{Q}}}),
\eqn
with $\Delta^2(k)=\Delta_{\text{GR}}^2(k) {\cal{Q}}$ and
\bqn
\Delta_{\text{GR}}^2(k)\equiv \frac{k^2}{8\pi^3 z^2} \Gamma^2(\nu) 2^{2\nu} y^{1-2\nu}.
\eqn
Then,  in Fig. \ref{fig4} we show that the relative errors of the power spectra can be dropped from $\sim 10\%$ to roughly $0.5\%$ for most of cases. 
However, from Fig.  \ref{fig4} we can also see  that in some very narrow regions of the parameter $b_2$, the relative errors become large. This usually happens when $g(y)$ has three real turning points and $\xi_0^2$ is very  large. When we derive the expression of ${\cal{Q}}$, the errors come from two parts. The first one is from the uniform approximation itself, while the other  is from the asymptotic expressions (\ref{Wasymp}) and (\ref{negative large}) that were used to determine the coefficients $\alpha_0$ and $\beta_0$. Both of them will lead to errors in the phase of $\cos{2\mathfrak{B}}$. We find that even with a very small such error, it could lead to a large error in the power spectra. To see this more explicitly, let us assume that $\mathfrak{B}$ has a small relative error $\delta$. When $\xi_0$ is large, the quantity ${\cal{A}}$ can be written as
\bqn
{\cal{A}}\simeq 2 e^{\pi \xi_0^2} \Big\{1+\cos{ \Big(2\mathfrak{B}+\delta\Big) }\Big\}.
\eqn
Now considering the region in which $\cos{2\mathfrak{B}}\simeq -1$,  while ${\cal{A}}$ is still very small. Then,  the resulted relative error for power spectra is about
\bqn
 \left|\frac{8 \cos{2\mathfrak{B}}}{1+\cos{2\mathfrak{B}}}\right| \delta \gg 1.
\eqn
The improvement of such errors requires the considerations  of   the more accurate expansions of the parabolic cylinder and Airy functions, and meanwhile   the high-order uniform approximations.
These are  out of the scope of this paper and will be considered  in our future work.

\section{Discussions and Conclusions}

In this paper, we have presented a technique, {\em the uniform asymptotic approximation}, to construct accurate analytical  solutions 
of  the linear  perturbations of inflation  with a nonlinear dispersion relation in order to account for the quantum effects of the early universe. 
Such a relation can be naturally realized in the framework of the Ho\v{r}ava-Lifshitz gravity  \cite{Horava,reviews}, a candidate of the 
ultraviolet complete theory of quantum gravity.

In particular, using the Liouville transformations of Eq.(\ref{Olver trans}),  we have first written the equation of motion of the mode function 
that describes the linear perturbations of scalar, vector or tensor fields in the form (\ref{eomU}) in terms of the two new variables $U(\xi)$ and $\xi$. 
According to  the theory of the second-order differential equations, the accuracy of the  approximate  solutions of  $U(\xi)$ sensitively 
depends on the behavior of the functions $g(y)$ and $q(y)$ defined in Eq.(\ref{gplusq}) around their poles and zeros (turning  points). 
To obtain the best approximate analytical solutions, we have first written them down formally in term of $g(y)$ and $q(y)$, and then constructed
the corresponding error bounds associated with the approximations. By minimizing the error bounds and error control functions, 
we have uniquely determined the two functions  $g(y)$ and $q(y)$, given by Eq.(\ref{function}). After obtaining all the individual solutions, using
the initial conditions, we have  then   uniquely determined the integrations constants, by matching them in their common regions. 
Because of the understanding of the error bounds, and the proper choice of the Liouville transformations near each pole or turning point,
our approximate analytical solutions  describe the  exact evolution of the linear 
perturbations  extremely well even only in the first-order approximations, as one can see from Figs. \ref{fig1} - \ref{fig4}.

In Section IV, we have considered a particular case  where   the parameter $|\xi_0|$ defined in 
Eqs.(\ref{xi0A}) and (\ref{xi0B})  is very large. When there is only one real root, this corresponds to the case considered in \cite{HL}. As shown explicitly in Appendix D, the solutions
constructed in this section can be also obtained from the ones constructed in Section III with the large  $|\xi_0|$ limit, as  it is expected.

As an application of the approximate analytical solutions constructed in Section III and IV, in Section V we have calculated the power spectra 
and indices of scalar and tensor perturbations  in the slow-roll inflation, and found that the amplitudes of the power spectra get modified 
due to the quantum effects, while the power spectrum indices remain the same as in the linear case, as far as the first-order approximations of the
slow-roll conditions are concerned. 

Certainly, the method developed in this paper can be easily applied to other cosmological backgrounds, and it would be very interesting to
study the quantum effects of the early universe on inflationary cosmology, including the power spectra and indices of the linear (scalar, vector and tensor)
perturbations,  non-Gaussianities,  primordial gravitational waves,  temperature and polarization of  CMB \cite{cosmoLV}.

In addition, high-order approximations can be also constructed \cite{Olver1974,Olver1975,Nayfeh}. 

\section*{Acknowledgements}

We thank Yongqing Huang, Jiro Soda,  Qiang Wu, and Wen Zhao for valuable discussions and suggestions.  This work is supported in part by DOE, DE-FG02-10ER41692 (AW), 
Ci\^encia Sem Fronteiras, No. 004/2013 - DRI/CAPES (AW), 
NSFC No. 11375153 (AW), No. 11173021 (AW), No. 11047008 (TZ), No. 11105120 (TZ), and No. 11205133 (TZ).

\section*{Appendix A: Theorems of Singular Integral Equations} 
\renewcommand{\theequation}{A.\arabic{equation}} \setcounter{equation}{0}

In this Appendix, we shall present a basic introduction to the theorems of the error bounds studied in this paper. 
For more details, we refer readers to \cite{Olver1974}.

Assume that $h(\xi)$ represents the errors of the approximate solutions of a second-order differential equation. One can substitute the 
corresponding solution with this error term to the corresponding  differential equation, and using the method of variation of parameters, 
one can cast $h(\xi)$ in the form,  
\bqn\lb{error}
h(\xi)&=&\int_{\alpha}^{\xi} {\cal{K}}(\xi,v) \Big[\phi(v)J(v)\nb\\
&&\;\;\;\;\;\;\;\;\;\;+\psi_0(\xi)h(v)+\psi_1(v)h'(v)\Big]dv,\nb\\
\eqn
where $J(v)$, $\phi(v)$, $\psi_0(v)$, and $\psi_1(v)$ are continuous functions, and the kennel ${\cal{K}}(\xi,v)$ and its first two partial $\xi$ derivatives are continuous. 

Then,  if the kennel ${\cal{K}}(\xi,v)$ and its first two $\xi$ derivatives are bounds, that is, if
\bqn
&&|{\cal{K}}(\xi,v)|\leq P_0(\xi) Q(v),\nb\\
&&\left|\frac{\partial {\cal{K}}(\xi,v) }{\partial \xi}\right|\leq P_1(\xi) Q(v),\nb\\
&&\left|\frac{\partial^2 {\cal{K}}(\xi,v) }{\partial \xi^2}\right| \leq P_2(\xi) Q(v),
\eqn
where the $P_j(\xi)$ and $Q(v)$ are continuous real functions and $P_j(\xi)$ are positive, we find
\bqn
\frac{|h(\xi)|}{P_0(\xi)},\;\frac{|h'(\xi)|}{P_1(\xi)} \leq \kappa \Phi(\xi) \exp{\{\kappa_0\Psi_0(\xi)+\kappa_1 \Psi_1(\xi)\}},\nb\\
\eqn
where 
\bqn
\Phi(\xi)&\equiv & \int_\alpha^\xi |\phi(v)|dv, \;\;\;
\Psi_0(\xi) \equiv\int_\alpha^\xi |\psi_0(v)|dv,\nb\\
\Psi_1(\xi) &\equiv& \int_\alpha^\xi |\psi_1(v)|dv,
\eqn
are  convergent, and   
\bqn
\kappa &\equiv& \text{sup} \{Q(\xi)|J(\xi)|\},\;\;\;
\kappa_0\equiv \text{sup} \{Q(\xi)P_0(\xi)\},\nb\\
\kappa_1 &\equiv& \text{sup} \{Q(\xi)P_1(\xi)\},
\eqn
are   finite.
Then, for the special case where $\psi_0(v)=\phi(v)$ and $\psi_1(v)=0$, the error bounds read \cite{Olver1974}
\bqn
\frac{|h(\xi)|}{P_0(\xi)},\;\frac{|h'(\xi)|}{P_1(\xi)} \leq \frac{\kappa}{\kappa_0}\Big [ \exp{\{\kappa_0\Psi_0(\xi)\}}-1\Big].
\eqn

\section*{Appendix B: Auxiliary Functions of Airy Functions}
\renewcommand{\theequation}{B.\arabic{equation}} \setcounter{equation}{0}

In this Appendix, we introduce some auxiliary functions of the  Airy functions $\text{Ai}(x)$ and $\text{Bi}(x)$. In particular, the
 modulus function $M(x)$ and $N(x)$ are defined as
\bqn
M(x)=\begin{cases}
\sqrt{\text{2Ai}(x) \text{Bi}(x)},\;& \text{for}\;x\leq c,\\
\sqrt{\text{Ai}^2(x)+\text{Bi}^2(x)}, & \text{for}\; x \geq c
\end{cases}\\
N(x)=\begin{cases}
\sqrt{\text{Ai}'^2(x)+\text{Bi}'^2(x)}\;& \text{for}\;x\leq c,\\
\sqrt{\frac{\text{Ai}'^2(x)\text{Bi}^2(x) +\text{Bi}'^2(x) \text{Ai}^2(x)}{\text{Ai}(x) \text{Bi}(x)}}, & \text{for}\; x \geq c,
\end{cases}
\eqn
where $c=-0.366046$ is the smallest root (in the sense of the absolute value) of the equation $\text{Ai}(x)=\text{Bi}(x)$. Then,  the weight function $E(x)$ is
defined as
\bqn
E(x)=\begin{cases}
\sqrt{\frac{\text{Bi}(x)}{\text{Ai(x)}}},&  \text{for}\;x\geq  c,\\
1, & \text{for}\; x \leq c.
\end{cases}
\eqn

With the phase functions $\theta(x)$ and $\omega(x)$,  the Airy functions can be expressed in terms of the modulus functions $M(x),\;N(x)$ and weight function $E(x)$ as
\bqn
\text{Ai}(x)=\frac{M(x)}{E(x)} \sin{\theta(x)},\;\;\text{Bi}(x)=E(x) M(x)\cos{\theta(x)},\nb\\
\text{Ai}'(x)=\frac{N(x)}{E(x)} \sin{\omega(x)},\;\;\text{Bi}'(x)=N(x) E(x)\cos{\omega(x)}.\nb\\
\eqn

\section*{Appendix C: Auxiliary Functions of Parabolic Cylinder Functions}
\renewcommand{\theequation}{C.\arabic{equation}} \setcounter{equation}{0}

In this Appendix, we present all the definitions of the auxiliary functions for the parabolic cylinder functions. First,
 we introduce the modulus functions $M\left(\frac{1}{2}\xi_0^2,\sqrt{2}\xi\right)$, $N\left(\frac{1}{2}\xi_0^2,\sqrt{2}\xi\right)$, 
 and the weight function $E\left(\frac{1}{2}\xi_0^2,\sqrt{2}\xi\right)$.  When $\sqrt{2}\xi\leq c(\xi_0)$, we have
\begin{widetext}
\bqn
&&M\left(\frac{1}{2}\xi_0^2,\sqrt{2}\xi\right)= \sqrt{2 W\left(\frac{1}{2}\xi_0^2,\sqrt{2}\xi\right) W\left(\frac{1}{2}\xi_0^2,-\sqrt{2}\xi\right)},\nb\\
&&N\left(\frac{1}{2}\xi_0^2,\sqrt{2}\xi\right)=\Bigg\{\frac{W'^2\left(\frac{1}{2}\xi_0^2,\sqrt{2}\xi\right) W\left(\frac{1}{2}\xi_0^2,
-\sqrt{2}\xi\right)}{W\left(\frac{1}{2}\xi_0^2,\sqrt{2}\xi\right)}-\frac{W'^2\left(\frac{1}{2}\xi_0^2,-\sqrt{2}\xi\right) W\left(\frac{1}{2}\xi_0^2,
\sqrt{2}\xi\right)}{W\left(\frac{1}{2}\xi_0^2,-\sqrt{2}\xi\right)} \Bigg\}^{1/2},
\eqn
and
\bqn
E\left(\frac{1}{2}\xi_0^2,\sqrt{2}\xi\right)=\sqrt{\frac{j(\xi_0) W\left(\frac{1}{2}\xi_0^2,-\sqrt{2}\xi\right)}{W\left(\frac{1}{2}\xi_0^2,\sqrt{2}\xi\right)}},
\eqn
where $E\left(\frac{1}{2}\xi_0^2,\sqrt{2}\xi\right)$ is of non-decreasing in  the region $\xi \in [0,+\infty)$,  and 
$c(\xi_0)$ is the smallest root, in the sense of the absolute value,  of the equation 
$$
j(\xi_0)^{-1/2} W'^2\left(\frac{1}{2}\xi_0^2,\sqrt{2}\xi\right)=j(\xi_0)^{1/2} W'^2\left(\frac{1}{2}\xi_0^2,-\sqrt{2}\xi\right).
$$
Then,  when $\sqrt{2} {x}\geq c(\xi_0)$, we have
\bqn
M\left(\frac{1}{2}\xi_0^2,\sqrt{2}\xi\right)&=& \Big\{j(\xi_0)^{-1}W^2\left(\frac{1}{2}\xi_0^2,\sqrt{2}\xi\right)+j(\xi_0) W^2\left(\frac{1}{2}\xi_0^2,-\sqrt{2}\xi\right)\Big\}^{1/2},\nb\\
N\left(\frac{1}{2}\xi_0^2,\sqrt{2}\xi\right)&=& \Big\{j(\xi_0)^{-1}W'^2\left(\frac{1}{2}\xi_0^2,\sqrt{2}\xi\right)+j(\xi_0) W'^2\left(\frac{1}{2}\xi_0^2,-\sqrt{2}\xi\right)\Big\}^{1/2},
\eqn
and
\bqn
E\left(\frac{1}{2}\xi_0^2,\sqrt{2}\xi\right)=1.
\eqn
Thus,  with the phase functions $\theta\left(\frac{1}{2}\xi_0^2,\sqrt{2}\xi\right)$ and $\omega\left(\frac{1}{2}\xi_0^2,\sqrt{2}\xi\right)$, the Parabolic cylinder functions 
can be expressed in terms of the modulus functions $M\left(\frac{1}{2}\xi_0^2,\sqrt{2}\xi\right),\;N\left(\frac{1}{2}\xi_0^2,\sqrt{2}\xi\right)$ and weight function 
$E\left(\frac{1}{2}\xi_0^2,\sqrt{2}\xi\right)$ as
\bqn
j(\xi_0)^{-1/2} W\left(\frac{1}{2}\xi_0^2,\sqrt{2}\xi\right) &=&\frac{M\left(\frac{1}{2}\xi_0^2,\sqrt{2}\xi\right)}{E\left(\frac{1}{2}\xi_0^2,\sqrt{2}\xi\right)} \sin{\theta\left(\frac{1}{2}\xi_0^2,\sqrt{2}\xi\right)},\nb\\
j(\xi_0)^{1/2} W\left(\frac{1}{2}\xi_0^2,-\sqrt{2}\xi\right) &=& M\left(\frac{1}{2}\xi_0^2,\sqrt{2}\xi\right) E\left(\frac{1}{2}\xi_0^2,\sqrt{2}\xi\right)\cos{\theta\left(\frac{1}{2}\xi_0^2,\sqrt{2}\xi\right)},\nb\\
j(\xi_0)^{-1/2} W'\left(\frac{1}{2}\xi_0^2,\sqrt{2}\xi\right) &=&\frac{N\left(\frac{1}{2}\xi_0^2,\sqrt{2}\xi\right)}{E\left(\frac{1}{2}\xi_0^2,\sqrt{2}\xi\right)} \sin{\omega\left(\frac{1}{2}\xi_0^2,\sqrt{2}\xi\right)},\nb\\
j(\xi_0)^{1/2} W'\left(\frac{1}{2}\xi_0^2,-\sqrt{2}\xi\right) &=& - N\left(\frac{1}{2}\xi_0^2,\sqrt{2}\xi\right) E\left(\frac{1}{2}\xi_0^2,\sqrt{2}\xi\right) \cos{\omega\left(\frac{1}{2}\xi_0^2,\sqrt{2}\xi\right)},
\eqn
for $\xi\geq 0$, and 
\bqn
j(\xi_0)^{1/2} W\left(\frac{1}{2}\xi_0^2,\sqrt{2}\xi\right) &=&M\left(\frac{1}{2}\xi_0^2,\sqrt{2}\xi\right) E\left(\frac{1}{2}\xi_0^2,\sqrt{2}\xi\right)\cos{\theta\left(\frac{1}{2}\xi_0^2,\sqrt{2}\xi\right)},\nb\\
j(\xi_0)^{-1/2} W\left(\frac{1}{2}\xi_0^2,-\sqrt{2}\xi\right) &=& \frac{M\left(\frac{1}{2}\xi_0^2,\sqrt{2}\xi\right)}{E\left(\frac{1}{2}\xi_0^2,\sqrt{2}\xi\right)} \sin{\theta\left(\frac{1}{2}\xi_0^2,\sqrt{2}\xi\right)},\nb\\
j(\xi_0)^{1/2} W'\left(\frac{1}{2}\xi_0^2,\sqrt{2}\xi\right) &=& - N\left(\frac{1}{2}\xi_0^2,\sqrt{2}\xi\right) E\left(\frac{1}{2}\xi_0^2,\sqrt{2}\xi\right) \cos{\omega\left(\frac{1}{2}\xi_0^2,\sqrt{2}\xi\right)},\nb\\
j(\xi_0)^{-1/2} W'\left(\frac{1}{2}\xi_0^2,-\sqrt{2}\xi\right) &=&  \frac{N\left(\frac{1}{2}\xi_0^2,\sqrt{2}\xi\right)}{E\left(\frac{1}{2}\xi_0^2,\sqrt{2}\xi\right)} \sin{\omega\left(\frac{1}{2}\xi_0^2,\sqrt{2}\xi\right)},
\eqn
for $\xi \leq 0$. 
\end{widetext}

\section*{Appendix D: Parabolic Cylinder Functions in Terms of Airy Functions}
\renewcommand{\theequation}{D.\arabic{equation}} \setcounter{equation}{0}

When $|\xi_0|$ is large, one can also express the parabolic cylinder functions in  terms of the Airy functions.

\subsubsection{When $y_{1,2}$ Are Real}

Near the real turning points $y_1$ and $y_2$,  the  parabolic cylinder functions for  $|\xi_0| \gg 1$ and $\xi >-\xi_0$ are given by
\bqn\lb{asyW}
W\left(\frac{1}{2}\xi_0^2,\sqrt{2}\xi\right) &\simeq& 2^{-1/4}\pi^{1/2} e^{-\pi \xi_0^2/4}\nb\\
&&\times \left(\frac{-\hat{\xi}}{\xi^2-\xi_0^2}\right)^{1/4} \text{Bi}(\hat{\xi}),\nb\\
W\left(\frac{1}{2}\xi_0^2,-\sqrt{2}\xi\right) &\simeq& 2^{3/4}\pi^{1/2} e^{\pi \xi_0^2/4} \nb\\
&& \times \left(\frac{-\hat{\xi}}{\xi^2-\xi_0^2}\right)^{1/4} \text{Ai}(\hat{\xi}),
\eqn
where in writing down the above expressions, we have neglected all the high-order terms and token 
\bqn\lb{hatxi1}
- \frac{2}{3}\hat{\xi}^{3/2}= \frac{1}{2}\xi \sqrt{\xi_0^2-\xi^2} -\frac{1}{2}\xi_0^2 \arccos{\left(-\frac{\xi}{\xi_0}\right)},
\eqn
for $\xi \in (-\xi_0,\xi_0)$, and
\bqn\lb{hatxi2}
 \frac{2}{3}\left(-\hat{\xi}\right)^{3/2}= \frac{1}{2}\xi \sqrt{\xi^2-\xi_0^2} -\frac{1}{2}\xi_0^2 \text{arcosh}\left(\frac{\xi}{\xi_0}\right),
\eqn
for $\xi\geq \xi_0$.

In the neighborhoods of $y_2$,  
we have  $\hat{\xi}=\xi_2$. Thus,  the solution (\ref{solutionW}) now takes the form
\bqn
\mu_k(y)&=& \alpha_1 2^{-1/4} \pi^{1/2} e^{-\pi \xi_0^2/4}  \left(\frac{\xi_2}{g(y)}\right)^{1/4} \text{Bi}(\xi_2)\nb\\
&&+\beta_1 2^{3/4} \pi^{1/2}e^{\pi \xi_0^2/4} \left(\frac{\xi_2}{g(y)}\right)^{1/4} \text{Ai}(\xi_2), ~~~
\eqn
which is exactly the asymptotic solution $\mu_k^{(2)}(y)$ given in Eq.(\ref{airy solution}) around the turning point $y_2$ with
\bqn\lb{coe1}
\hat{\beta}_2&=&\alpha_1 2^{-1/4} \pi^{1/2} e^{-\pi \xi_0^2/4},\nb\\
\hat{\alpha}_2&=&\beta_1 2^{3/4} \pi^{1/2} e^{\pi \xi_0^2/4}.
\eqn

Considering the solutions in the neighborhoods of $y_1$, similarly one can find now $\hat{\xi}=\xi_1$. Then,  taking the asymptotic 
expansions (\ref{asyW}) of the parabolic functions into account, we find that   the solution (\ref{solutionW}) now takes the form
\bqn
\mu_k(y)&=& \alpha_1 2^{3/4} \pi^{1/2} e^{\pi \xi_0^2/4} \left(\frac{\xi_1}{g(y)}\right)^{1/4} \text{Ai}(\xi_1)\nb\\
&&+\beta_1 2^{-1/4} \pi^{1/2} e^{-\pi \xi_0^2/4} \left(\frac{\xi_1}{g(y)}\right)^{1/4} \text{Bi}(\xi_1),\nb\\
\eqn
which is the same  as the asymptotic solution $\mu_k^{(1)}(y)$ given in Eq.(\ref{airy solution}) around the turning point $y_1$,  with
\bqn \lb{coe2}
\hat{\alpha}_1&=& \alpha_1 2^{3/4} \pi^{1/2} e^{\pi \xi_0^2/4},\nb\\
\hat{\beta}_1&=& \beta_1 2^{-1/4} \pi^{1/2} e^{-\pi\xi_0^2/4}.
\eqn
With $\alpha_1$ and $\beta_1$ given by Eq.(\ref{ab}), it can be shown that  $\hat{\alpha}_2,\; \hat{\beta}_2$ given in Eq.(\ref{coe1}) 
and $\hat{\alpha}_1,\; \hat{\beta}_1$ given  in Eq.(\ref{coe2}) are consistent with Eqs.(\ref{ab1}) and (\ref{ab2}),  except an irrelevant 
phase factor.

\subsubsection{When $y_1$ and $y_2$ Are Complex}

For complex $y_1$ and $y_2$, when $|\xi_0|$ is large, we find that $\frac{1}{2}\xi_0^2$ is very negative. Then,  as mentioned above, the  solutions given in term of the 
Airy functions  in Eq.(\ref{airy solution1}) are valid in the whole region $y\in (0^+,\infty)$, so the solutions in this case are already written in terms   of the 
Airy functions, and no further considerations are needed.

\section*{Appendix E: The Elliptic Integrals}
\renewcommand{\theequation}{E.\arabic{equation}} \setcounter{equation}{0}

In this section, we  present various  definitions of the Elliptic integrals used in this paper. For detail, see, for example,  \cite{AS72}. 
First, $F(\phi,m)$ denotes  the Elliptic integral of the first kind, defined as  
\bqn
F(\phi,m) =\int_0^\phi \frac{d\theta}{\sqrt{1-m \sin^2{\theta}}},
\eqn
for  $-\frac{\pi}{2}<\phi\leq\frac{\pi}{2}$. 
$K(m)$ denotes  the complete Elliptic integral of the first kind,  defined as
\bqn
K(m)=F\left(\frac{\pi}{2},m\right).
\eqn

The Elliptic integral $E(\phi,m)$ of the second kind  is defined as, 
\bqn
E(\phi,m) =\int_0^\phi \sqrt{1-m \sin^2{\theta}}d\theta,
\eqn
for  $-\frac{\pi}{2}<\phi\leq\frac{\pi}{2}$, while the corresponding complete Elliptic integral $E(m)$ of the second kind  is defined as
\bqn
E(m)=E\left(\frac{\pi}{2},m\right).
\eqn

$\Pi(n,\phi,m)$ denotes the Elliptic integral of the third kind, defined as
\bqn
\Pi(n,\phi,m)=\int_0^\phi \frac{d\theta}{(1-n \sin^2{\theta})\sqrt{1-m \sin^2{\theta}}},
\eqn
and the corresponding complete Elliptic integral $\Pi(n,m)$ is defined as
\bqn
\Pi(n,m)=\Pi\left(n,\frac{\pi}{2},m\right).
\eqn

\end{document}